# On pseudorapidity distribution and speed of sound in high energy heavy ion collisions based on a new revised Landau hydrodynamic model

Li-Na Gao and Fu-Hu Liu[1]

*Institute of Theoretical Physics, Shanxi University, Taiyuan, Shanxi 030006, China*

**Abstract**

We propose a new revised Landau hydrodynamic model to study systematically the pseudorapidity distributions of charged particles produced in heavy ion collisions over an energy range from a few GeV to a few TeV per nucleon pair. The interacting system is divided into three sources namely the central, target, and projectile sources respectively. The large central source is described by the Landau hydrodynamic model and further revised by the contributions of the small target/projectile sources. In the calculation, to avoid the errors caused by an unapt conversion or non-division, the rapidity and pseudorapidity distributions are obtained respectively. The modeling results are in agreement with the available experimental data at relativistic heavy ion collider (RHIC), large hadron collider (LHC), and other energies for different centralities. The value of square speed of sound parameter in different collisions has been extracted by us from the widths of rapidity distributions. Our results show that, in heavy ion collisions at RHIC and LHC energies, the central source undergoes through a phase transition from hadronic gas to quark-gluon plasma (QGP) liquid phase; meanwhile, the target/projectile sources remain in the state of hadronic gas. The present work confirms that the QGP is of liquid type rather than that of a gas. The whole region of participants undergoes through a mixed phase consisting of a large quantity of (>90%) QGP liquid and a small quantity of (<10%) hadronic gas.

## 1. Introduction

In fields of particle physics and nuclear physics, heavy ion (nucleus-nucleus) collisions at high energies are a very important research subject. Many charged and neutral particles are produced in final state of the collisions and can be measured in experiments. To study the behavior of the particles could help us to understand the processes of interacting system in the collisions. The pseudorapidity distributions of charged particles are an important quantity which can be measured in the early stage of the measurements. According to ecumenical textbooks, the pseudorapidity $\eta$ is simply defined as $\eta = -\ln\tan(\theta/2)$, where $\theta$ is the emission angle of the considered particle.

The pseudorapidity distributions can be used to study stopping and penetrating powers of the target and projectile nuclei, positions and contribution ratios of different emission sources, contribution ratios of leading nucleons, square speed of sound, and other related topics. Many models have been introduced in order to study the pseudorapidity distributions, transverse momentum distributions, azimuthal correlations, and other distributions and correlations in relativistic heavy ion collisions. These theoretical models can be classified mainly into two classes: i) thermal and statistical model, and ii) transport and dynamical model. Among them, the three-fireball model [1-6], the three-source relativistic diffusion model [7-12], and the Landau hydrodynamic model [13-21] are of great interested for us and will be used in the present work.

---

[1] E-mail: fuhuliu@163.com; fuhuliu@sxu.edu.cn

A lot of experimental data on nuclear collisions at high energies has been published in literature. The relativistic heavy ion collider (RHIC) has performed gold-gold (Au-Au), copper-copper (Cu-Cu), deuteron-gold (*d*-Au), and other collisions at various GeV energies [22-25]. Also, the large hadron collider (LHC) has performed lead-lead (Pb-Pb), proton-lead (*p*-Pb), and other collisions at TeV energies [26]. These collider experiments show rich and exciting results on the pseudorapidity distributions and other distributions. In fixed target experiments as well (such as in nuclear emulsion experiments at high energies), proton to gold induced emulsion (*p*-Em and Au-Em) collisions have presented pseudorapidity distributions with abundant structures [27-29]. More other ions such as helium, carbon, oxygen, neon, silicon, sulphur, krypton, and etc. have also been used [30, 31].

In this paper, we propose a new revision on the Landau hydrodynamic model based on the three-source picture to describe systematically the pseudorapidity distributions of charged particles produced in Au-Au, Cu-Cu, Pb-Pb, *d*-Au, Au-Em, and *p*-Em collisions at high energies. The central source with a large enough contribution is described by the Landau hydrodynamic model. The small contributions of the target and projectile sources are considered as a revision on the Landau hydrodynamic model for the central source. Based on the descriptions of pseudorapidity distributions, the values of square speed of sound parameter are extracted from the widths of rapidity distributions.

## 2. The model and calculation method

Enlightening by the three-fireball model [1-6] and the three-source relativistic diffusion model [7-12], we classify the particle emission sources into three types: a central source (C), a target source (T), and a projectile source (P). Generally, the central source's center stays at the mid-rapidity, and its contribution covers a wide enough region in the whole rapidity distribution. The central source includes the contributions of all produced particles and most non-leading nucleons. The target source stays in the left side of the central source's center and covers an appropriate region. It includes the contributions of all leading and a few non-leading target nucleons. The projectile source stays in the right side of the central source's center and covers an appropriate region. It includes the contributions of all leading and a few non-leading projectile nucleons. As revisions of the central source, the distribution ranges of particles produced from the target and projectile sources are covered by that from the central source.

The central source can be described by the Landau Hydrodynamic model [13-21]. The target and projectile sources are assumed to emit isotropically particles in their respective rest fames. The three sources may have different contribution ratios which are regarded as free parameters in the model. The central source can produce pions, kaons, nucleons, and other particles. The target and projectile sources emit only nucleons due to the two sources consisting of leading and non-leading nucleons. In fact, the target/projectile sources are a complementarity and revision of the central source. Generally, spectator nucleons appear in the very backward or forward rapidity region in non-central collisions; their contributions should be subtracted in our analyses. Most experimental distributions don't contain the contributions of spectator nucleons due to relative central collisions and narrow region of measurement. Our treatment is in fact a new revision of the Landau Hydrodynamic model [13-21].

In center-of-mass reference frame and for symmetric collisions such as Au-Au and Pb-Pb collisions, the central source stays at the peak position at $y_C = 0$, the target source stays at a peak position in the range of $y_T < 0$, and the projectile source stays at a peak position in the range of $y_P > 0$. In an actual calculation for symmetric collisions, $y_T$ and $y_P$ are regarded as free parameters. The distributions contributed by the target/projectile sources revise (in fact increase) somewhere the probabilities underestimated by the central source. In the case of considering asymmetric collisions such as $d$-Au collisions, the central source stays at a peak position at $y_C < 0$, and the situations for the target/projectile sources are similar to those in symmetric collisions. In an actual calculation for asymmetric collisions, $y_C$, $y_T$, and $y_P$ are regarded as free parameters, where $y_C$ is close to the peak position. In the fixed target experiments in laboratory reference system, $y_C > 0$ for both symmetric and asymmetric collisions. Obviously, we have $y_T < y_C < y_P$ in any case.

In center-of-mass or laboratory reference frame, according to the Landau hydrodynamic model [13-21], particles produced in the central source can be described by a Gaussian rapidity ($y$) distribution with a width of $\sigma$ [15-21]. We have

$$\frac{dN}{dy} = \frac{N_0}{\sqrt{2\pi}\sigma} \exp\left[-\frac{(y - y_C)^2}{2\sigma^2}\right], \tag{1}$$

where $N$ denotes the multiplicity, $N_0$ is the normalization constant, and $\sigma$ should be large enough to cover a wide enough rapidity region. At the same time, the transverse momentum ($p_T$) distribution is assumed to obey the simplest form of Boltzmann distribution [32]

$$f_{p_T}(p_T) = C_0 p_T \exp\left(-\frac{\sqrt{p_T^2 + m_0^2}}{kT}\right), \tag{2}$$

where $C_0$ is the normalization, $m_0$ denotes the rest mass of the considered particle, $k$ denotes the Boltzmann constant, and $T$ denotes the source temperature. In the simplest form [Eq. (2)], the chemical potential and the distinction for fermions and bosons are not included due to small effects at high energy.

Based on Eqs. (1) and (2), we can get a series of values of pseudorapidity for the particles produced from the central source by using the Monte Carlo method. Then, we can get the pseudorapidity distribution contributed by the central source by the statistical method. Now we describe the process which calculates one of values of pseudorapidity. Let $R_{1,2,3}$ denote random numbers in [0,1]. The Gaussian rapidity distribution [Eq. (1)] results in $y$ to be

$$y = \sigma\sqrt{-2\ln R_1}\cos(2\pi R_2) + y_C. \qquad (3)$$

A given $p_T$ satisfies

$$\int_0^{p_T} f_{p_T}(p_T)dp_T < R_3 < \int_0^{p_T+dp_T} f_{p_T}(p_T)dp_T. \qquad (4)$$

The longitudinal momentum $p_z$ can be given by

$$p_z = \sqrt{p_T^2 + m_0^2}\sinh y. \qquad (5)$$

The momentum $p$ is

$$p = \sqrt{p_T^2 + p_z^2}. \qquad (6)$$

Then, we have the pseudorapidity $\eta$ to be

$$\eta = \frac{1}{2}\ln\left(\frac{p+p_z}{p-p_z}\right). \qquad (7)$$

Repeating the above process [Eqs. (3)-(7)] for many times, a series of values of pseudorapidity for the particles produced in the central source can be obtained.

The situation for the target/projectile sources is somewhat different from that for the central source. We now describe the process which calculates one of values of pseudorapidity for the particles produced from the target or projectile source. Let $R_4$ denote random number in [0,1]. In the rest frame of the target or projectile source, particles are assumed to emit isotropically. The Monte Carlo method gives the emission angle $\theta'$ to be

$$\theta' = \arctan\left[\frac{2\sqrt{R_4(1-R_4)}}{1-2R_4}\right] + \theta_0, \qquad (8)$$

where $\theta_0 = 0$ (or $\pi$) in the case of the first item in the above equation being greater than 0 (or less than 0). This isotropic emission results in a Gaussian pseudorapidity distribution with the width of 0.91-0.92 [27]. In this study, we have not studied further the width corresponding to the target and projectile sources. The transverse momentum $p_T$ has the same expression as Eqs. (2) and (4). The longitudinal momentum $p_z'$, the momentum $p'$, and the energy $E'$ are written by

$$p_z' = p_T \cot\theta', \qquad (9)$$

$$p' = \sqrt{p_T^2 + p_z'^2}, \qquad (10)$$

and

$$E' = \sqrt{p'^2 + m_0^2}, \qquad (11)$$

respectively. In laboratory or center-of-mass reference frame, the rapidity $y$ of the particle produced in the target or projectile source can be given by

$$y = \frac{1}{2}\ln\left(\frac{E'+p_z'}{E'-p_z'}\right) + y_{T,P}. \qquad (12)$$

The longitudinal momentum $p_z$, the momentum $p$, and the pseudorapidity $\eta$ have the same expressions as Eqs. (5)-(7) respectively. Repeating the above process for many times, we can obtain a series of values of $\eta$ for the particles produced in the target or projectile source.

If necessary, $\eta$ distribution for particles produced in the central source can be obtained by the statistical method. At the same time, $\eta$ distribution for particles produced in the target or projectile source can be obtained by the statistical method, too. To compare with experimental $\eta$ distribution, we have to use the statistical method to collect lots of $\eta$ for particles produced in the central, target, and projectile sources. In the statistical process, the contribution ratios of the three sources have to be considered naturally. A reasonable and accordant comparison with experimental data can determine a set of parameters. Particularly, the width $\sigma$ of rapidity distribution for particles produced in the central source is the most interesting parameter, although other parameters can be obtained together.

The relation between the square speed of sound $c_s^2$ and the rapidity distribution width $\sigma$ can be given by [15, 16, 18, 20, 21]

$$\sigma = \sqrt{\frac{8}{3}\frac{c_s^2}{1-c_s^4}\ln\left(\frac{\sqrt{s_{NN}}}{2m_N}\right)}, \tag{13}$$

where $m_N$ denotes the mass of a proton. Then, $c_s^2$ is expressed by using $\sigma$ to be

$$c_s^2 = \frac{1}{3\sigma^2}\left[\sqrt{16\ln^2\left(\frac{\sqrt{s_{NN}}}{2m_N}\right)+9\sigma^4} - 4\ln\left(\frac{\sqrt{s_{NN}}}{2m_N}\right)\right]. \tag{14}$$

The square speed of sound parameter can be obtained from the above equation. It should be noted that not only the central source but also the target/projectile sources discussed in the present work are formed in the participants, but not in the spectators which are mainly appeared in non-central or peripheral collisions. The relation between the square speed of sound and rapidity distribution width can be applied for the groups of particles produced from the three sources.

We would like to point out that, as the previous version, it applies the hydrodynamic model for hadron production that was originally developed by Landau [13], and Belen'kji and Landau [14]. Subsequently other researchers have extended it [15-20], always based on an expanding central source for the produced particles in the rapidity/pseudorapidity space. The variance of this Gaussian source in Landau's original work is in the pseudorapidity space and later works in the rapidity space. Generally, the variance of this Gaussian source is related analytically to the logarithm of the center-of-mass energy, and the speed of sound. Hence the speed of sound can in principle be inferred from a comparison of calculated (pseudo-)rapidity distributions with data given this particular model. We think that the application of Gaussian source in the rapidity space is a revision of that in the pseudorapidity space. Therefore, the Gaussian rapidity distribution is used in the present work.

In addition, according to the Landau hydrodynamic model [13-21], in hadron-nucleus and nucleus-nucleus collisions, the number and energy of primary hadrons, which form an ideal liquid, fluctuate from event to event. Therefore, in already formed ideal liquid in laboratory system, the (pseudo)rapidity of its center-of-mass also fluctuates on the (pseudo)rapidity scale. The total inclusive distribution is then the sum of Gaussian distributions with different centers and widths, and the shape of the inclusive distributions differs from the form of a Gaussian distribution, in the laboratory system. These situations are changed in the center-of-mass system, where the mentioned centers are the same, and the mentioned widths fluctuate slightly from event to event within a given centrality range. Because our calculation is performed in the center-of-mass system and for the given centrality ranges in most cases, we will not take into account the fluctuations of the center and width of Gaussian distribution for the central source from event to event for the purpose of convenience. Even if for a few mini-bias samples and emulsion experiments in the laboratory system, our treatment gives an average for different events. For the target and projectile sources in their respective rest frames, we will not take into account the fluctuations of the center either, and the width is fixed due to isotropic emission and given temperature.

In the calculation, we don't need either to consider the Jacobian transformation between the rapidity and pseudorapidity spaces explicitly, but instead calculate pseudorapidity distribution for produced charged hadrons directly in the Monte Carlo approach, in which we assume Gaussian rapidity distribution and Boltzmann transverse momentum distribution for the central sources, and isotropic emission and Boltzmann transverse momentum distribution for the target and projectile sources in their respective rest frames. At the same time, we assume Landau's prescription for the speed of sound to be valid for the three sources. Based on the description of experimental pseudorapidity distribution, speed of sound parameters are then determined for the three sources from their respective widths of rapidity distributions.

3. **Comparisons with experimental data**

The pseudorapidity distributions, $dN_{ch}/d\eta$, of charged particles produced in Au-Au collisions for different centralities at $\sqrt{s_{NN}} = 19.6$, 62.4, 130, and 200 GeV are presented in Figures 1-4, respectively. In the four figures, the circles represent the experimental data measured by the PHOBOS collaboration [22], and the curves are our results calculated by the Monte Carlo method, where the spectator contributions in Figures 1(e)-1(j) are not considered. In the calculation, we take $m_0 = 0.174$ GeV/$c^2$ for the particles produced in the central source and $m_0 = 0.938$ GeV/$c^2$ for the particles produced in the target and projectile sources respectively. The former one is estimated by us from an average weighed the masses and yields of $\pi^\pm$, $K^\pm$, $\bar{p}$, and $p$ [33]. The later one is the mass of a proton. The temperature is taken to be 0.156 GeV [34]. The values of peak position $y_T$ and contribution ratio $k_T$ of the target source, the rapidity distribution width $\sigma$ of the central source, as well as $\chi^2$ per degree of freedom ($\chi^2/dof$) obtained in the fitting are given in Table 1, where the contributions of spectators in Figures 1(e)-1(j) are not counted in the description, which result in larger $\chi^2/dof$ if we consider them in the calculation of $\chi^2/dof$, and there is no spectator to be expected in other figures. The last two columns in Table 1 will be discussed later. To estimate the values of parameters, the least-squared fitting method is used. For the symmetric collisions, the peak position $y_P$ and contribution ratio $k_P$ of the projectile source, and the peak position $y_C$ and contribution ratio $k_C$ of the central source, can be given by $y_P = -y_T$, $k_P = k_T$,

$y_C = 0$, and $k_C = 1 - k_T - k_P$, respectively.

One can see from Figures 1-4 that, in the case of subtracting the spectator nucleons, the modeling results describe the experimental data. In the considered energy and centrality ranges, $y_T$ has a left shift with increase of the energy $\sqrt{s_{NN}}$ and does not shown an obvious change with the centrality percentage $C$, $k_T$ does not show an obvious change with $\sqrt{s_{NN}}$ and $C$, $\sigma$ increases with increase of $\sqrt{s_{NN}}$ and $C$. Because nuclear penetrating power increases with increase of $\sqrt{s_{NN}}$, the dependence trends of $y_T$ and $\sigma$ on $\sqrt{s_{NN}}$ are naturally. This also renders the dependence of $\sigma$ on $C$ due to larger penetrating power appearing in peripheral collisions. In Figures 1(e)-1(j), to give a better description, the contributions of spectator nucleons have to be considered. Although this energy (19.6 GeV) is close to the region of critical energy from the hadronic matter to quark-gluon plasma (QGP) [35], we focus our attention on the produced particles in the present work, but not the spectator nucleons.

Figures 5-7 are similar to those in Figures 1-4, but they show the results in Cu-Cu collisions with different centralities at $\sqrt{s_{NN}} = 22.4$, 62.4, and 200 GeV, respectively, where the spectator contributions in Figures 5(d)-5(l) are not considered in the calculation. The values of $y_T$, $k_T$, $\sigma$, and $\chi^2/dof$ obtained in the fitting are given in Table 2, where the contributions of spectators in Figures 5(d)-5(l) are not counted in the description, which result in larger $\chi^2/dof$ if we consider them in the calculation of $\chi^2/dof$, and there is no spectator to be expected in other figures. The last two columns in Table 2 will be discussed later. Likewise, one can see that the modeling results describe the experimental data of the PHOBOS collaboration [22]. In the considered energy and centrality ranges, the dependence trends of parameters on $\sqrt{s_{NN}}$ and $C$ are the same as those obtained from Figures 1-4 and Table 1.

The pseudorapidity distributions of charged particles produced in Pb-Pb collisions for different centralities at $\sqrt{s_{NN}} = 2.76$ TeV are presented in Figure 8. The circles represent the experimental data measured by ALICE collaboration [26], and the curves represent the results calculated by us. The values of $y_T$, $k_T$, $\sigma$, and $\chi^2/dof$ obtained in the fitting are given in Table 3. The last two columns in Table 3 will be discussed later. One can see that the results calculated by us are in agreement with the experimental data. All of the three free parameters have no obvious dependence on $\sqrt{s_{NN}}$ and $C$ in the error range and in the centrality range of 0-30%. Because $k_T$ in Table 3 is very small ($\leq 0.3\%$), the contributions of target and projectile sources can be neglected.

The pseudorapidity distributions of charged particles produced in $d$-Au collisions for minimum-bias sample and different centralities at $\sqrt{s_{NN}} = 0.2$ TeV are shown in Figures 9(a) and 9(b)-9(f) respectively. The circles represent the experimental data quoted in references [23, 24] [Figure 9(a)] and [25] [Figures 9(b)-9(f)], and the curves are the results calculated by us. The values of peak positions ($y_T$, $y_C$, $y_P$), contribution ratios ($k_T$, $k_P$), width $\sigma$, and $\chi^2/dof$ obtained in the fitting are given in Table 4. The last two columns in Table 4 will be discussed later. One can see that the results calculated by us are in agreement with the experimental data. Because the considered collisions are asymmetric, we have to use more parameters. From Table 4 one can see that $y_T$ and $k_T$ have almost no change, $y_C$ and $y_P$ have right shifts, $k_P$ increases, and $\sigma$ decreases with increase of $C$. These results are caused by the very asymmetric collisions. It should be noted that *d*-Au collisions have been analyzed by Wolschin et al. in their previous work [11] within their three-source model [7-12]. The present work is to some extent a repetition of the original work of Wolschin et al. [11], although different methods are used.

Figure 10 shows the pseudorapidity distributions, $(1/N_{EV})dN/dy$, of shower (relativistic singly charged) particles produced in Au-Em collisions with different impacting types at beam energy $E_{\text{beam}} = 10.6\text{-}10.7A$ GeV, where $N$ and $N_{EV}$ denotes the numbers of shower particles and events respectively. The circles represent the experimental data quoted in reference [27], and the curves are the results calculated by us. The impacting types such as very central, semi-central, peripheral, and very peripheral are assumed by us to correspond to centrality percentages 10%, 40%, 60%, and 90%, respectively. The values of $y_T$, $y_C$, $y_P$, $k_T$, $k_P$, $\sigma$, and $\chi^2/dof$ are listed in Table 5. The last two columns in Table 5 will be discussed later. One can see that the experimental data can be described by our modeling results. The parameters $y_T$, $k_T$, and $k_P$ don't show an obvious change, the parameters $y_C$ and $y_P$ show right shifts, and the parameter $\sigma$ shows an increase when the centrality percentage increases.

The pseudorapidity distributions of combined shower and grey (cascaded singly charged) particles produced in minimum-bias *p*-Em collisions at $\sqrt{s_{NN}} = 6.7$, 11.2, 19.4, 23.7, and 38.7 GeV are presented in Figures 11(a)-11(e), respectively. The circles represent the experimental data quoted in references [28, 29], and the curves are our modeling results. By fitting the experimental data of the fixed target emulsion experiments, we get the values of $y_T$, $y_C$, $y_P$, $k_T$, $k_P$, $\sigma$, and $\chi^2/dof$ listed in Table 6. The last two columns in Table 6 will be discussed later. The modeling results describe the experimental data. It is shown that $y_T$ doesn't change obviously, $y_C$, $y_P$, $k_T$, and $\sigma$ increase, and $k_P$ decreases with increase of $\sqrt{s_{NN}}$.

To see clearly the dependences of peak positions ($y_T$, $y_C$, $y_P$) and width $\sigma$ on centrality percentage $C$ or energy $\sqrt{s_{NN}}$, we plot the parameter values listed in Tables 1-6 in Figures 12 and 13 respectively, where all the values of points are extracted from the modeling results when we compare the model with experimental data. The symbols and lines represent the parameter values and linear fitting respectively. For symmetric collisions, only $y_T - C$ relations are presented due to $y_C - C$ and $y_P - C$ relations to be trivial. All information on energies and types are marked in the figure panels. The intercepts, slopes, and $\chi^2/dof$ corresponding to the lines in Figures 12 and 13 are listed in Tables 7 and 8 respectively. The conclusions obtained from Figures 1-11 and Tables 1-6 can be seen clearly in Figures 12 and 13. Other dependences such as the dependences of contribution ratios ($k_T$, $k_P$) on $C$ or $\sqrt{s_{NN}}$ are not plotted due to trivialness of these relations.

According to the widths of rapidity distributions, the extracted values of square speed of sound parameter for different interacting systems, centrality percentages, and energies are listed in the last two columns in Tables 1-6. To see clearly the dependences of $c_s^2$ [for the target (or projectile) and central sources] on $C$ or $\sqrt{s_{NN}}$, we plot the values listed in the last two columns in Tables 1-6 in Figures 14(a)-14(g) or 14(h), where all the values of points are extracted from the modeling results when we compare the model with experimental data. The symbols and lines represent the parameter values and linear fitting respectively. Only $c_s^2 - C$ or $c_s^2 - \sqrt{s_{NN}}$ relations for the target and central sources are presented due to these relations for the projectile source to be trivial. All information on the energies and types are marked in the figure panels. The intercepts, slopes, and $\chi^2/dof$ corresponding to the lines in Figure 14 are listed in Table 9. One can see that, in most cases, $c_s^2$ for the central source is greater than that for the target source, which renders that the central source has more density due to more energy deposit. The values of $c_s^2(C)$ for central Au-Au and Cu-Cu collisions are less than those for peripheral collisions, which renders that the central collisions undergoes through longer transverse expanding time which results in lower density due to larger volume.

Although we studied the square speed of sound parameter in our previous work [36, 37], the physics picture and calculation method used there are different from the present work. In our previous work, we used a four-source picture, two (target and projectile) participants and two (target and projectile) spectators. The rapidity and pseudorapidity distributions were not distinguished for the purpose of convenience. In addition, our previous work seems not to correspond with the three-source picture [1-12]. The present work uses a large enough central source which is described by the Landau hydrodynamic model [13-21] and two small (target and projectile) sources which revise somewhere the contribution of the central source. At the same time, the rapidity and pseudorapidity distributions are strictly distinguished, which avoids the errors caused by non-division or inapposite conversion. Comparing with the results of the four-source picture in our previous work, larger $c_s^2(C)$ values are obtained by the new revised Landau Hydrodynamic model in the present work.

If the three-source picture is correct, the present results on larger $c_s^2(C)$ ($\approx 0.5$) imply an important experimental phenomenon. According to the relation $c_s^2 = 1/D$ which is for zero shear modulus and massless particles, where $D$ denotes the dimensionality of space [38, 39], we know that $c_s^2 = 1/3$ for the ideal hadronic gas [40-49] which has $D$=3 in the reaction volume and $c_s^2 = 1/2$ [39] for $D$=2 which corresponds to the ideal QGP liquid phase in the reaction plane. The present work shows that the target/projectile sources remain in the state of hadronic gas in all cases and the central source undergoes through the phase transition from the hadronic gas to the QGP liquid in most cases. At the same time, $c_s^2(C)$ for central collisions is less than that for peripheral collisions, which renders a transformation from the state of reaction volume in central collisions to the state of reaction plane in peripheral collisions. In the error ranges and in most cases, we have $1/3 < c_s^2(C) \leq 1/2$ in collisions at RHIC and LHC energies. A few $c_s^2(C)$ are greater than 0.5 due to the statistical fluctuations.

## 4. Discussions

Although we have used the four-source picture in our previous work [36, 37] which tries to combine the Landau hydrodynamic model [13-21] with the Glauber model [50] for participants and spectators and ends up with four Gaussian sources for particle production, it doesn't mean that a mid-rapidity source is completely missing. In fact, the separation for target and projectile participants give us only a convenience to consider the contributions of target and projectile nuclei in collisions. The sub-sources in the target participant are mainly contributed by the target nucleus, and the sub-sources in the projectile participant are mainly contributed by the projectile nucleus. The separation for target and projectile participants doesn't mean that there is no sub-source in the mid-rapidity region. Instead, we can regard together the target and projectile participants as whole of the three fireballs or sources in the three-fireball [1-6] or three-source model [7-12].

In our opinion, the central source discussed in the revised Landau hydrodynamic model in the present work contains almost all of the target and projectile participants, which results in a wide enough Gaussian rapidity distribution. The target and projectile sources contain only a few leading and non-leading nucleons, which revise the Gaussian rapidity distribution somewhere. The related picture in the revised Landau hydrodynamic model is set up at the level of nucleon, which is different from, but not inharmonious to, the three-fireball model [1-6] and the three-source model [7-12], which separate the whole participants into three parts: the central (mid-rapidity), target, and projectile sources.

Obviously, the central (mid-rapidity) fireball in the three-fireball model [1-6] or the central (mid-rapidity) source in the three-source model [7-12] is a part of the central source in the present work. Similarly, the target/projectile fireballs (sources) in the former two models are parts of the central source in the present work. Meanwhile, in the present work, the target/projectile sources are revisions of the central source. Particularly, the physics picture of three-source model [7-12] is set up at the level of quark and gluon. According to the three-source model [7-12], in the approximate energy range of $\sqrt{s_{NN}} > 20$ GeV, the low-$x$ gluon density rises rapidly with center-of-mass energy, and particles produced from the mid-rapidity source are most relevant at RHIC energies and dominant at LHC energies; while in the approximate energy range of $\sqrt{s_{NN}} < 20$ GeV, no mid-rapidity source occurs [12]. These results contain more underlying and embedded physics such as interactions among quarks and gluons. We are interested in details of the three-source model [7-12], and hope to use this model in our future work.

From above calculations, comparisons, and discussions we see that, based on a new revised Landau hydrodynamic model, we present a new parameterization of pseudorapidity distributions of charged particles produced in heavy ion collisions for different energies and centralities. This parameterization is not baseless. From this parameterization, we have obtained a series of square speed of sound, particularly some of them being $1/3 < c_s^2(C) \leq 1/2$. These results are novel and interesting. We have explained $1/3 < c_s^2(C) \leq 1/2$ in heavy ion collisions at RHIC and LHC energies that the central source undergoes through the phase transition from the hadronic gas to the QGP liquid. This confirms the QGP is of liquid type rather than of a gas.

## 5. Conclusions

From the above discussions, we obtain following conclusions:

(a) The interacting system formed in high energy heavy ion collisions can be divided into three sources: a central source, a target source, and a projectile source. The central source is large enough covering the whole rapidity region and can be described by the Landau hydrodynamic model. The small target and projectile sources which isotropically emit particles in their respective rest frames are used revising the rapidity or pseudorapidity distributions described by the Landau hydrodynamic model. All the three sources are assumed to be formed in the participants. Our treatment on the sources is a new revision of the Landau hydrodynamic model, and this revision is different from previous revisions. The three-source picture used in the present work is also different from previous three-fireball or three-source model.

(b) The rapidity and pseudorapidity distributions are strictly dipartite in the modeling calculation. There is no conversion being used between the two distributions, where an unapt conversion may cause some errors. At the same time, a non-division for the two distributions may also cause some errors. The modeling results calculated by the Monte Carlo method are in agreement with the available experimental data of Au-Au, Cu-Cu, and $d$-Au collisions at the RHIC, Pb-Pb collisions at the LHC, as well as Au-Em and $p$-Em collisions in the fixed target emulsion experiments. The considered collisions cover an energy range from a few GeV to a few TeV. Although we have done the parameterization of pseudorapidity distributions, this treatment is not baseless. In fact, it is based on the new revised Landau hydrodynamic model. Based on the parameterization, we have extracted some useful and interesting information, particularly the square speed of sound.

(c) Our results calculated by the new revised Landau hydrodynamic model shows that, in most cases, in collider experiments, $y_T$ has a left shift with an increase of the energy $\sqrt{s_{NN}}$ and does not show an obvious change with the centrality percentage $C$, $k_T$ does not show an obvious change with $\sqrt{s_{NN}}$ and $C$, and $\sigma$ increases with an increase of $\sqrt{s_{NN}}$ and $C$. In the fixed target experiments, $y_T$, $k_T$, and $k_P$ don't show an obvious change, $y_C$ and $y_P$ show right shifts, and $\sigma$ shows an increase when the centrality percentage increases; In addition, $y_T$ doesn't show an obvious change, $y_C$, $y_P$, $k_T$, and $\sigma$ increase, and $k_P$ decreases with increase of $\sqrt{s_{NN}}$. Although these behaviors of parameters are not too extraordinary, they are obtained from the new consideration and based on many experimental data.

(d) From the widths of rapidity distributions, the values of square speed of sound in different collisions are obtained. Because more energy deposit and then more density are reached in the central source, the central source corresponds to larger $c_s^2$ than the target/projectile sources. At the same time, comparing with the peripheral collisions, the central collisions undergoes through longer transverse expanding time which results in lower density due to larger volume. Comparing with our previous work, the new revised Landau Hydrodynamic model obtains a larger $c_s^2$ for the central source. Although it is hard to say that the present work is better, it supplies a new consideration for us to extract the square speed of sound.

(e) In the framework of the three-source picture, in the error ranges and in most cases, we have $1/3 < c_s^2(C) \leq 1/2$ in heavy ion collisions at RHIC and LHC energies. These values of $c_s^2(C)$ implies that the central source undergoes through the phase transition from the hadronic gas to the QGP liquid. At the same time, the target/projectile sources stay in the state of hadronic gas in all cases due to $c_s^2(T) < 1/3$. The present work confirms that the QGP is of liquid type rather than that of a gas. The whole region of participants undergoes through a mixed phase consisting of a large quantity of (>90%) QGP liquid and a small quantity of (<10%) hadronic gas. These highlighted results are obtained by us from the extraction of square speed of sound from rapidity distribution width based on the parameterization of pseudorapidity distributions in the framework of new revised Landau Hydrodynamic model.

**Conflict of Interests**

The authors declare that there is no conflict of interests regarding the publication of this paper.

**Acknowledgments**

This work was supported by the National Natural Science Foundation of China under Grant no. 10975095, the Open Research Subject of the Chinese Academy of Sciences Large-Scale Scientific Facility under Grant no. 2060205, the Shanxi Provincial Natural Science Foundation under Grant No. 2013021006, and the Shanxi Scholarship Council of China under Grant no. 2012-012.

**References**

[1] A. D'innocenzo, G. Ingrosso, and P. Rotelli, "A new two-component model for $pp$ topological cross-sections," *Il Nuovo Cimento A*, vol. 44, pp. 375–391, 1978.

[2] A. D'innocenzo, G. Ingrosso, and P. Rotelli, Lett. "The $p\bar{p}$ annihilation channel as a prototype for the central fireball in $pp$ production processes," *Lettere Nuovo Cimento*, vol. 25, pp. 393–398, 1979.

[3] A. D'innocenzo, G. Ingrosso, and P. Rotelli, "A universal scaling function for hadron-hadron interactions," *Il Nuovo Cimento A*, vol. 55, pp. 417–466, 1980.

[4] A. D'innocenzo, G. Ingrosso, and P. Rotelli, "The three-component fireball model and $p\bar{p}$ interactions," *Lettere al Nuovo Cimento*, vol. 27, pp. 457–466, 1980.

[5] L.-S. Liu and T.-C. Meng, "Multiplicity and energy distributions in high-energy $e^+e^-$, $pp$, and $p\bar{p}$ collisions," *Physical Review D*, vol. 27, pp. 2640–2647, 1983.

[6] K.-C. Chou, L.-S. Liu, and T.-C. Meng, "Koba-Nielsen-Olesen scaling and production mechanism in high-energy collisions," *Physical Review D*, vol. 28, pp. 1080–1085, 1983.

[7] G. Wolschin, "Relativistic diffusion model," *The European Physical Journal A*, vol. 5, pp. 85–90, 1999.

[8] G. Wolschin, "Diffusion in relativistic systems," *Progress in Particle and Nuclear Physics*, vol. 59, pp. 374–382, 2007.

[9] G. Wolschin, "Pseudorapidity distributions of produced charged hadrons in pp collisions at RHIC and LHC energies," *EPL*, vol. 95, Article ID 61001, 2011.

[10] G. Wolschin, "Particle production sources at LHC energies," *Journal of Physics G*, vol 40, Article ID 045104, 2013.

[11] G. Wolschin, M. Biyajima, T. Mizoguchi, and N. Suzuki, "Local thermalization in the d + Au system," *Physics Letters B*, vol. 633, pp. 38–42, 2006.

[12] G. Wolschin, "Ultraviolet energy dependence of particle production sources in relativistic heavy-ion collisions," *Physical Review C*, vol. 91, Article ID 014905, 2015.

[13] L. D. Landau, "Izvestiya akademii nauk: series fizicheskikh 17 51 (1953)," in *English-Translation: Collected Papers of L. D. Landau*, D. Ter-Haarp, Ed., p. 569, Pergamon, Oxford, UK, 1965.

[14] S. Z. Belen'kji and L. D. Landau, "Hydrodynamic theory of multiple production of particles," *Il Nuovo Cimento*, vol. 3, no. 1 supplement, pp. 15–31, 1956.

[15] E. V. Shuryak, "Multiparticle production in high energy particle collisions,"

*Yadernaya Fizika*, vol. 16, no. 2, pp. 395–405, 1972.
[16] O. V. Zhirov and E. V. Shuryak, "Multiple production of hadrons and predictions of the Landau theory," *Yadernaya Fizika*, vol. 21, pp. 861–867, 1975.
[17] P. Carruthers and M. Duong-Van, "New scaling law based on the hydrodynamical model of particle production," *Physics Letters B*, vol. 41, no. 5, pp. 597–601, 1972.
[18] P. Carruthers and M. Duong-Van, "Rapidity and angular distributions of charged secondaries according to the hydrodynamical model of particle production," *Physical Review D*, vol. 8, pp. 859–874, 1973.
[19] P. Steinberg, "Landau hydrodynamics and RHIC phenomena," *Acta Phys. Hung. A*, vol. 24, pp. 51–57, 2005
[20] C.-Y. Wong, "Landau hydrodynamics reexamined," *Physical Review C*, vol. 78, Article ID 054902, 2008.
[21] M. Gazdzicki, M. Gorenstein, and P. Seyboth, "Onset of deconfinement in nucleus-nucleus collisions: review for pedestrians and experts," *Acta Physica Polonica B*, vol. 42, pp. 307–351, 2011.
[22] B. Alver, B. B. Back, D. S. Barton et al., "Charged-particle multiplicity and pseudorapidity distributions measured with the PHOBOS detector in Au+Au, Cu+Cu, d+Au, and p+p collisions at ultrarelativistic energies," *Physical ReviewC*, vol. 83, Article ID 024913, 2011.
[23] B. B. Back, M. D. Baker, M. Ballintijn et al., "Pseudorapidity distribution of charged particles in d+Au collisions at $\sqrt{s_{NN}} = 200$ GeV," *Physical Review Letters*, vol. 93, Article ID 082301, 2004.
[24] I. Arsene et al., BRAHMS Collaboration, "Centrality dependence of charged-particle pseudorapidity distributions from d+Au collisions at $\sqrt{s_{NN}} = 200$ GeV," *Physical Review Letters*, vol. 94, Article ID 032301, 2005.
[25] B. B. Back, M. D. Baker, M. Ballintijn et al., "Scaling of charged particle production in d+Au collisions at $\sqrt{s_{NN}} = 200$ GeV," *Physical Review C*, vol. 72, Article ID 031901, 2005.
[26] E. Abbas, L. Aphecetche, A. Baldisseri et al., "Centrality dependence of the pseudorapidity density distribution for charged particles in Pb-Pb collisions at $\sqrt{s_{NN}} = 2.76$ TeV," *Physics Letters B*, vol. 726, pp. 610-622, 2013.
[27] M. I. Adamovich et al., EMU01 Collaboration, "Charged particle density distributions in Au induced interactions with emulsion nuclei at 10.7A GeV," *Physics Letters B*, vol. 352, pp. 472–478, 1995.
[28] A. Abduzhamilov, L. Barbier, L. P. Chernova et al., "Charged-particle multiplicity and angular distributions in proton-emulsion interactions at 800 GeV." *Physical Review D*, vol. 35, pp. 3537–3540, 1987.
[29] I. Otterlund, E. Stenlund, B. Andersson et al., "Nuclear interactions of 400 GeV protons in emulsion", *Nuclear Physics B*, vol. 142, pp.445–462, 1978.
[30] M. A. Rahim, S. Fakhraddin, and H. Asharabi, "Systematic study of projectile fragments in nucleus-nucleus collisions at 4.1-4.5 A GeV/c and multi-source thermal model," *The European Physical Journal A*, vol. 48, Article ID 115, 2012.
[31] M. A. Rahim and S. Fakhraddin, "Study of events in absence of target/projectile fragments in interactions of nuclei with emulsion at 4.5A GeV/c and 4.1A GeV/c," *Nuclear Physics A*, vol. 831, pp. 39–48, 2009.

[32] C.-R. Meng, "Transverse momentum and rapidity distributions of φ mesons produced in Pb-Pb collisions at SPS energies," *Chinese Physics Letters*, vol. 26, Article ID 102501, 2009.

[33] A. Adare et al., PHENIX Collaboration, "Spectra and ratios of identified particles in Au+Au and d+Au collisions at $\sqrt{s_{NN}}$ =200 GeV," *Physical Review C*, vol. 88, Article ID 024906, 2013.

[34] P. Braun-Munzinger, "Hadron production in nuclear collisions and the QCD phase boundary," Talk given at 45 years of nuclear theory at Stony Brook: a tribute to Gerald E. Brown, Stony Brook University, Stony Brook, USA, 2013, http://tonic.physics.sunysb.edu/gerrybrown/program2.html.

[35] S. Chatterjee, S. Das, L. Kumar, D. Mishra, B. Mohanty, R. Sahoo, and N. Sharma, "Freeze out parameters in heavy-ion collisions at AGS, SPS, RHIC and LHC Energies," *Advances in High Energy Physics*, vol. 2015, Article ID 349013, 2015.

[36] L.-N. Gao, Y.-H. Chen, H.-R. Wei, and F.-H. Liu, "Speed of sound parameter from RHIC and LHC heavy-ion data," *Advances in High Energy Physics*, vol. 2013, Article ID 450247, 2013.

[37] Y.-Q. Gao, T. Tian, L.-N. Gao, and F.-H. Liu, "Pseudorapidity distribution of charged particles and square speed of sound parameter in $p$-$p$ or $p$-$\bar{p}$ collisions over an energy range from 0.053 to 7 TeV," *Advances in High Energy Physics*, vol. 2014, Article ID 569079, 2014.

[38] K.-Y. Kim and I. Zahed, "Baryonic response of dense holographic QCD," *Journal of High Energy Physics*, vol. 0812, Article ID 075, 2008.

[39] E. I. Buchbinder, A. Buchel, and S. E. Vázquez, "Sound Waves in (2+1) Dimensional Holographic Magnetic Fluids," *Journal of High Energy Physics*, vol. 0812, Article ID 090, 2008.

[40] A. Bazavov, T. Bhattacharya, C. DeTar et al., "The equation of state in (2+1)-flavor QCD," arXiv: 1407.6387v2.

[41] U. Heinz, P. Sorensen, A. Deshpande et al., "Exploring the properties of the phases of QCD matter – research opportunities and priorities for the next decade," arXiv: 1501.06477v2.

[42] B. Betz, "Jet propagation and Mach-cone formation in (3+1)-dimensional ideal hydrodynamics," arXiv: 0910.4114, Ph.D. Thesis, University Frankfurt, Germany.

[43] P. Castorina, J. Cleymans, D. E. Miller, and H. Satz, "The speed of sound in hadronic matter," *The European Physical Journal C*, vol. 66, pp. 207–213, 2010.

[44] V. Roy and A. K. Chaudhuri, "Equation of state dependence of Mach cone like structures in Au+Au collisions," *Journal of Physics G*, vol. 37, Article ID 035105, 2010.

[45] A. Cherman and T. D. Cohen, "A bound on the speed of sound from holography," *Physical Review D*, vol. 80, Article ID 066003, 2009.

[46] P. M. Hohler and M. A. Stephanov, "Holography and the speed of sound at high temperatures," *Physical Review D*, vol. 80, Article ID 066002, 2009.

[47] C. P. Herzog and S. S. Pufu, "The second sound of SU(2)," *Journal of High Energy Physics*, vol. 0904, Article ID 126, 2009.

[48] S. K. Ghosh, T. K. Mukherjee, M. G. Mustafa, and R. Ray, "QGP susceptibilities from PNJL model," *Indian Journal of Physics*, vol. 85, pp. 87–91, 2011.

[49] U. Gürsoy, E. Kiritsis, L. Mazzanti, and F. Nitti, "Deconfinement and gluon

plasma dynamics in improved holographic QCD," *Physical Review Letters*, vol. 101, Article ID 181601, 2008.

[50] R. J. Glauber, "High-energy collision theory," in *Lectures of Theoretical Physics*, W. E. Brittin and L. G. Dunham, Eds., vol. 1, pp. 315–414, Interscience, New York, NY, USA, 1959.

Table 1. Values of peak position $y_T$ and contribution ratio $k_T$ of the target source, the rapidity distribution width $\sigma$ of the central source, as well as $\chi^2/dof$ for the curves in Figures 1-4 which show Au-Au collisions for different energies and centralities. For the symmetric collisions, the peak position $y_P$ and contribution ratio $k_P$ of the projectile source and the peak position $y_C$ and contribution ratio $k_C$ of the central source can be given by $y_P = -y_T$, $k_P = k_T$, $y_C = 0$, and $k_C = 1 - k_T - k_P$, respectively. The last two columns show the values of square speed of sound for the target [$c_s^2(T)$] or projectile [$c_s^2(P) = c_s^2(T)$] and central sources [$c_s^2(C)$] respectively.

| Figure | Bin (%) | $y_T$ | $k_T$ | $\sigma$ | $\chi^2/dof$ | $c_s^2(T)$ | $c_s^2(C)$ |
|---|---|---|---|---|---|---|---|
| 19.6 GeV | | | | | | | |
| Figure 1(a) | 0-3 | -1.20±0.25 | 0.03±0.01 | 1.60±0.08 | 0.413 | 0.130±0.028 | 0.357±0.028 |
| Figure 1(b) | 3-6 | -1.10±0.25 | 0.02±0.01 | 1.65±0.08 | 0.603 | 0.130±0.028 | 0.374±0.027 |
| Figure 1(c) | 6-10 | -1.10±0.25 | 0.02±0.01 | 1.70±0.08 | 0.198 | 0.130±0.028 | 0.391±0.027 |
| Figure 1(d) | 10-15 | -1.00±0.25 | 0.02±0.01 | 1.75±0.08 | 0.325 | 0.130±0.028 | 0.408±0.027 |
| Figure 1(e) | 15-20 | -1.07±0.20 | 0.01±0.01 | 1.80±0.08 | 0.984 | 0.130±0.028 | 0.424±0.026 |
| Figure 1(f) | 20-25 | -1.02±0.20 | 0.01±0.01 | 1.85±0.10 | 2.709 | 0.130±0.028 | 0.441±0.032 |
| Figure 1(g) | 25-30 | -1.00±0.20 | 0.01±0.01 | 1.85±0.10 | 6.506 | 0.130±0.028 | 0.441±0.032 |
| Figure 1(h) | 30-35 | -1.00±0.18 | 0.01±0.01 | 1.90±0.10 | 8.764 | 0.130±0.028 | 0.456±0.032 |
| Figure 1(i) | 35-40 | -1.04±0.20 | 0.02±0.02 | 1.90±0.06 | 11.531 | 0.130±0.028 | 0.457±0.019 |
| Figure 1(j) | 40-45 | -1.00±0.20 | 0.01±0.01 | 2.00±0.10 | 9.690 | 0.130±0.028 | 0.487±0.030 |
| 62.4 GeV | | | | | | | |
| Figure 2(a) | 0-3 | -1.70±0.21 | 0.04±0.01 | 2.00±0.09 | 0.411 | 0.088±0.019 | 0.370±0.025 |
| Figure 2(b) | 3-6 | -1.70±0.21 | 0.04±0.01 | 2.00±0.09 | 0.788 | 0.088±0.019 | 0.370±0.025 |
| Figure 2(c) | 6-10 | -1.60±0.20 | 0.04±0.01 | 2.00±0.10 | 0.522 | 0.088±0.019 | 0.370±0.028 |
| Figure 2(d) | 10-15 | -1.60±0.20 | 0.04±0.01 | 2.10±0.09 | 0.216 | 0.088±0.019 | 0.397±0.025 |
| Figure 2(e) | 15-20 | -1.80±0.20 | 0.04±0.01 | 2.15±0.11 | 0.247 | 0.088±0.019 | 0.411±0.030 |
| Figure 2(f) | 20-25 | -1.80±0.19 | 0.03±0.01 | 2.25±0.10 | 0.077 | 0.088±0.019 | 0.438±0.026 |
| Figure 2(g) | 25-30 | -1.70±0.20 | 0.03±0.01 | 2.30±0.11 | 0.117 | 0.088±0.019 | 0.451±0.029 |
| Figure 2(h) | 30-35 | -1.60±0.29 | 0.02±0.01 | 2.40±0.10 | 0.090 | 0.088±0.019 | 0.476±0.025 |
| Figure 2(i) | 35-40 | -1.60±0.30 | 0.01±0.01 | 2.50±0.09 | 0.116 | 0.088±0.019 | 0.501±0.022 |
| Figure 2(j) | 40-45 | -1.60±0.28 | 0.01±0.01 | 2.55±0.12 | 0.178 | 0.088±0.019 | 0.513±0.026 |
| Figure 2(k) | 45-50 | -1.70±0.20 | 0.01±0.01 | 2.55±0.12 | 0.208 | 0.088±0.019 | 0.513±0.028 |
| 130 GeV | | | | | | | |
| Figure 3(a) | 0-3 | -1.90±0.28 | 0.04±0.02 | 2.30±0.10 | 0.533 | 0.073±0.016 | 0.395±0.025 |
| Figure 3(b) | 3-6 | -1.90±0.30 | 0.04±0.02 | 2.30±0.09 | 0.423 | 0.073±0.016 | 0.395±0.023 |
| Figure 3(c) | 6-10 | -2.00±0.29 | 0.04±0.01 | 2.33±0.09 | 0.309 | 0.073±0.016 | 0.402±0.022 |
| Figure 3(d) | 10-15 | -2.00±0.30 | 0.04±0.01 | 2.40±0.07 | 0.193 | 0.073±0.016 | 0.420±0.017 |
| Figure 3(e) | 15-20 | -2.10±0.35 | 0.04±0.01 | 2.40±0.10 | 0.084 | 0.073±0.016 | 0.420±0.024 |

| Figure | Bin (%) | $y_T$ | $k_T$ | $\sigma$ | $\chi^2/dof$ | $c_s^2(T)$ | $c_s^2(C)$ |
|---|---|---|---|---|---|---|---|
| Figure 3(f) | 20-25 | -2.10±0.35 | 0.04±0.01 | 2.40±0.11 | 0.117 | 0.073±0.016 | 0.420±0.027 |
| Figure 3(g) | 25-30 | -2.10±0.32 | 0.03±0.01 | 2.50±0.10 | 0.128 | 0.073±0.016 | 0.444±0.024 |
| Figure 3(h) | 30-35 | -2.15±0.33 | 0.03±0.01 | 2.55±0.10 | 0.104 | 0.073±0.016 | 0.456±0.023 |
| Figure 3(i) | 35-40 | -2.15±0.30 | 0.03±0.01 | 2.60±0.10 | 0.057 | 0.073±0.016 | 0.467±0.023 |
| Figure 3(j) | 40-45 | -2.15±0.30 | 0.02±0.01 | 2.70±0.10 | 0.237 | 0.073±0.016 | 0.490±0.022 |
| Figure 3(k) | 45-50 | -2.15±0.30 | 0.02±0.01 | 2.70±0.12 | 0.096 | 0.073±0.016 | 0.490±0.027 |
| 200 GeV | | | | | | | |
| Figure 4(a) | 0-3 | -2.00±0.38 | 0.04±0.01 | 2.40±0.12 | 0.276 | 0.066±0.014 | 0.392±0.029 |
| Figure 4(b) | 3-6 | -2.00±0.49 | 0.04±0.01 | 2.40±0.13 | 0.251 | 0.066±0.014 | 0.392±0.031 |
| Figure 4(c) | 6-10 | -2.00±0.49 | 0.03±0.01 | 2.50±0.15 | 0.214 | 0.066±0.014 | 0.415±0.035 |
| Figure 4(d) | 10-15 | -2.10±0.42 | 0.03±0.01 | 2.60±0.15 | 0.144 | 0.066±0.014 | 0.438±0.034 |
| Figure 4(e) | 15-20 | -2.10±0.48 | 0.03±0.01 | 2.60±0.13 | 0.082 | 0.066±0.014 | 0.438±0.030 |
| Figure 4(f) | 20-25 | -2.10±0.48 | 0.03±0.01 | 2.60±0.15 | 0.108 | 0.066±0.014 | 0.438±0.034 |
| Figure 4(g) | 25-30 | -2.10±0.48 | 0.03±0.01 | 2.70±0.11 | 0.149 | 0.066±0.014 | 0.461±0.024 |
| Figure 4(h) | 30-35 | -2.10±0.50 | 0.03±0.01 | 2.70±0.15 | 0.096 | 0.066±0.014 | 0.461±0.033 |
| Figure 4(i) | 35-40 | -2.10±0.53 | 0.03±0.01 | 2.70±0.15 | 0.141 | 0.066±0.014 | 0.461±0.033 |
| Figure 4(j) | 40-45 | -2.10±0.50 | 0.03±0.01 | 2.70±0.15 | 0.188 | 0.066±0.014 | 0.461±0.033 |
| Figure 4(k) | 45-50 | -2.10±0.50 | 0.03±0.01 | 2.80±0.13 | 0.119 | 0.066±0.014 | 0.483±0.028 |

Table 2. As for Table 1, but showing the results for the curves in Figures 5-7 which show Cu-Cu collisions for different energies and centralities.

| Figure | Bin (%) | $y_T$ | $k_T$ | $\sigma$ | $\chi^2/dof$ | $c_s^2(T)$ | $c_s^2(C)$ |
|---|---|---|---|---|---|---|---|
| 22.4 GeV | | | | | | | |
| Figure 5(a) | 0-3 | -1.20±0.20 | 0.03±0.02 | 1.60±0.08 | 0.238 | 0.123±0.026 | 0.342±0.027 |
| Figure 5(b) | 3-6 | -1.10±0.19 | 0.01±0.01 | 1.70±0.08 | 0.115 | 0.123±0.026 | 0.375±0.027 |
| Figure 5(c) | 6-10 | -1.10±0.17 | 0.01±0.01 | 1.78±0.10 | 0.130 | 0.123±0.026 | 0.402±0.033 |
| Figure 5(d) | 10-15 | -1.12±0.20 | 0.01±0.01 | 1.82±0.11 | 1.264 | 0.123±0.026 | 0.415±0.035 |
| Figure 5(e) | 15-20 | -1.14±0.23 | 0.01±0.01 | 1.85±0.06 | 2.796 | 0.123±0.026 | 0.424±0.019 |
| Figure 5(f) | 20-25 | -1.10±0.21 | 0.01±0.01 | 1.94±0.06 | 3.001 | 0.123±0.026 | 0.453±0.019 |
| Figure 5(g) | 25-30 | -1.10±0.20 | 0.01±0.01 | 1.95±0.07 | 4.056 | 0.123±0.026 | 0.456±0.022 |
| Figure 5(h) | 30-35 | -1.04±0.20 | 0.01±0.01 | 1.96±0.10 | 6.165 | 0.123±0.026 | 0.459±0.031 |
| Figure 5(i) | 35-40 | -1.07±0.17 | 0.01±0.01 | 1.98±0.10 | 8.068 | 0.123±0.026 | 0.465±0.030 |
| Figure 5(j) | 40-45 | -1.02±0.20 | 0.01±0.01 | 2.02±0.08 | 10.256 | 0.123±0.026 | 0.477±0.024 |
| Figure 5(k) | 45-50 | -1.05±0.20 | 0.01±0.01 | 2.07±0.08 | 8.595 | 0.123±0.026 | 0.491±0.023 |
| Figure 5(l) | 50-55 | -1.10±0.22 | 0.01±0.01 | 2.10±0.08 | 13.713 | 0.123±0.026 | 0.500±0.023 |
| 62.4 GeV | | | | | | | |
| Figure 6(a) | 0-3 | -1.80±0.21 | 0.05±0.01 | 2.00±0.07 | 0.264 | 0.088±0.019 | 0.370±0.020 |
| Figure 6(b) | 3-6 | -1.50±0.21 | 0.04±0.01 | 2.00±0.09 | 0.342 | 0.088±0.019 | 0.370±0.025 |
| Figure 6(c) | 6-10 | -1.60±0.20 | 0.04±0.01 | 2.10±0.10 | 0.262 | 0.088±0.019 | 0.370±0.028 |
| Figure 6(d) | 10-15 | -1.60±0.20 | 0.04±0.01 | 2.13±0.10 | 0.170 | 0.088±0.019 | 0.406±0.027 |
| Figure 6(e) | 15-20 | -1.40±0.28 | 0.03±0.01 | 2.18±0.08 | 0.100 | 0.088±0.019 | 0.419±0.022 |
| Figure 6(f) | 20-25 | -1.40±0.30 | 0.03±0.01 | 2.25±0.10 | 0.096 | 0.088±0.019 | 0.438±0.026 |
| Figure 6(g) | 25-30 | -1.40±0.28 | 0.02±0.01 | 2.30±0.08 | 0.133 | 0.088±0.019 | 0.451±0.021 |
| Figure 6(h) | 30-35 | -1.40±0.30 | 0.02±0.01 | 2.35±0.11 | 0.129 | 0.088±0.019 | 0.464±0.028 |

| Figure | Bin (%) | $y_T$ | $k_T$ | $\sigma$ | $\chi^2/dof$ | $c_s^2(T)$ | $c_s^2(C)$ |
|---|---|---|---|---|---|---|---|
| Figure 6(i) | 35-40 | -1.40±0.25 | 0.02±0.01 | 2.45±0.10 | 0.123 | 0.088±0.019 | 0.489±0.024 |
| Figure 6(j) | 40-45 | -1.40±0.18 | 0.01±0.01 | 2.48±0.10 | 0.095 | 0.088±0.019 | 0.496±0.024 |
| Figure 6(k) | 45-50 | -1.40±0.20 | 0.01±0.01 | 2.48±0.08 | 0.159 | 0.088±0.019 | 0.496±0.019 |
| Figure 6(l) | 50-55 | -1.40±0.20 | 0.01±0.01 | 2.55±0.10 | 0.176 | 0.088±0.019 | 0.513±0.024 |
| 200 GeV | | | | | | | |
| Figure 7(a) | 0-3 | -2.35±0.28 | 0.04±0.02 | 2.50±0.08 | 0.194 | 0.066±0.014 | 0.415±0.019 |
| Figure 7(b) | 3-6 | -2.35±0.28 | 0.04±0.02 | 2.55±0.10 | 0.114 | 0.066±0.014 | 0.427±0.023 |
| Figure 7(c) | 6-10 | -2.30±0.25 | 0.04±0.02 | 2.60±0.10 | 0.085 | 0.066±0.014 | 0.438±0.023 |
| Figure 7(d) | 10-15 | -2.25±0.24 | 0.03±0.01 | 2.65±0.12 | 0.125 | 0.066±0.014 | 0.450±0.027 |
| Figure 7(e) | 15-20 | -2.30±0.25 | 0.04±0.02 | 2.65±0.12 | 0.121 | 0.066±0.014 | 0.450±0.027 |
| Figure 7(f) | 20-25 | -2.25±0.27 | 0.03±0.01 | 2.70±0.08 | 0.129 | 0.066±0.014 | 0.461±0.018 |
| Figure 7(g) | 25-30 | -2.25±0.32 | 0.03±0.02 | 2.78±0.08 | 0.076 | 0.066±0.014 | 0.479±0.017 |
| Figure 7(h) | 30-35 | -2.25±0.35 | 0.03±0.02 | 2.78±0.10 | 0.082 | 0.066±0.014 | 0.479±0.022 |
| Figure 7(i) | 35-40 | -2.25±0.33 | 0.03±0.02 | 2.85±0.10 | 0.053 | 0.066±0.014 | 0.494±0.021 |
| Figure 7(j) | 40-45 | -2.25±0.35 | 0.03±0.02 | 2.85±0.08 | 0.074 | 0.066±0.014 | 0.494±0.017 |
| Figure 7(k) | 45-50 | -2.25±0.32 | 0.03±0.02 | 2.85±0.10 | 0.083 | 0.066±0.014 | 0.494±0.021 |
| Figure 7(l) | 50-55 | -2.25±0.32 | 0.03±0.02 | 2.95±0.07 | 0.085 | 0.066±0.014 | 0.514±0.014 |

Table 3. As for Table 1, but showing the results for the curves in Figure 8 which shows Pb-Pb collisions at 2.76 TeV with different centralities. The values of $k_T$ are very small and can be neglected in the treatment.

| Figure | Bin (%) | $y_T$ | $k_T$ | $\sigma$ | $\chi^2/dof$ | $c_s^2(T)$ | $c_s^2(C)$ |
|---|---|---|---|---|---|---|---|
| Figure 8(a) | 0-5 | -0.50±0.06 | 0.0030±0.0005 | 3.60±0.05 | 0.069 | 0.042±0.009 | 0.500±0.008 |
| Figure 8(b) | 5-10 | -0.50±0.05 | 0.0018±0.0003 | 3.62±0.03 | 0.089 | 0.042±0.009 | 0.503±0.005 |
| Figure 8(c) | 10-20 | -0.40±0.05 | 0.0010±0.0003 | 3.62±0.08 | 0.114 | 0.042±0.009 | 0.504±0.013 |
| Figure 8(d) | 20-30 | -0.50±0.06 | 0.0001±0.0001 | 3.62±0.08 | 0.132 | 0.042±0.009 | 0.504±0.013 |

Table 4. Values of peak positions ($y_T$, $y_P$, $y_C$), contribution ratios ($k_T$, $k_P$), wide $\sigma$, and $\chi^2/dof$ for the curves in Figure 9 which shows $d$-Au collisions at 0.2 TeV for minimum-bias sample [Figure 9(a)] or with different centralities [Figures 9(b)-9(f)].

| Figure | Bin (%) | $y_T$ | $y_C$ | $y_P$ | $k_T$ | $k_P$ | $\sigma$ | $\chi^2/dof$ | $c_s^2(T)$ | $c_s^2(C)$ |
|---|---|---|---|---|---|---|---|---|---|---|
| Figure 9(a) | Mini Bias | -3.90±0.60 | -1.17±0.25 | 1.10±0.20 | 0.01±0.01 | 0.02±0.01 | 2.80±0.03 | 0.030 | 0.066±0.014 | 0.483±0.006 |
| Figure 9(b) | 0-20 | -4.70±0.41 | -1.70±0.23 | 0.60±0.30 | 0.02±0.01 | 0.02±0.01 | 2.90±0.07 | 0.096 | 0.066±0.014 | 0.504±0.014 |
| Figure 9(c) | 20-40 | -4.00±0.23 | -1.10±0.23 | 1.00±0.20 | 0.02±0.01 | 0.02±0.01 | 2.90±0.07 | 0.067 | 0.066±0.014 | 0.504±0.014 |
| Figure 9(d) | 40-60 | -4.10±0.18 | -0.80±0.22 | 1.10±0.22 | 0.02±0.01 | 0.02±0.01 | 2.80±0.07 | 0.031 | 0.066±0.014 | 0.483±0.015 |
| Figure 9(e) | 60-80 | -4.00±0.20 | -0.50±0.25 | 1.20±0.22 | 0.01±0.01 | 0.03±0.01 | 2.75±0.07 | 0.100 | 0.066±0.014 | 0.472±0.015 |
| Figure 9(f) | 80-100 | -4.20±0.20 | -0.50±0.23 | 1.35±0.25 | 0.02±0.01 | 0.09±0.01 | 2.20±0.07 | 0.334 | 0.066±0.014 | 0.343±0.017 |

Table 5. As for Table 4, but showing the results for the curves in Figure 10 which shows Au-Em collisions at $E_{\text{beam}}=10.6A$ GeV with different impacting types.

| Figure | Type | $y_T$ | $y_C$ | $y_P$ | $k_T$ | $k_P$ | $\sigma$ | $\chi^2/dof$ | $c_s^2(T)$ | $c_s^2(C)$ |
|---|---|---|---|---|---|---|---|---|---|---|
| Figure 10(a) | Central | 0.90±0.20 | 1.90±0.17 | 2.50±0.13 | 0.10±0.04 | 0.22±0.03 | 0.75±0.05 | 0.841 | 0.310±0.056 | 0.222±0.027 |
| Figure 10(b) | Semi-central | 0.95±0.05 | 2.40±0.10 | 3.70±0.10 | 0.20±0.04 | 0.17±0.02 | 1.00±0.05 | 1.522 | 0.310±0.056 | 0.360±0.028 |
| Figure 10(c) | Peripheral | 0.95±0.05 | 3.50±0.15 | 4.00±0.10 | 0.26±0.02 | 0.28±0.02 | 1.00±0.10 | 2.207 | 0.310±0.056 | 0.360±0.056 |
| Figure 10(d) | Very peripheral | 1.00±0.10 | 3.00±0.10 | 4.00±0.10 | 0.14±0.04 | 0.43±0.03 | 1.10±0.20 | 2.154 | 0.310±0.056 | 0.415±0.107 |

Table 6. As for Table 4, but showing the results for the curves in Figure 11 which shows $p$-Em collisions at different $\sqrt{s_{NN}}$.

| Figure | $\sqrt{s_{NN}}$ | $y_T$ | $y_C$ | $y_P$ | $k_T$ | $k_P$ | $\sigma$ | $\chi^2/dof$ | $c_s^2(T)$ | $c_s^2(C)$ |
|---|---|---|---|---|---|---|---|---|---|---|
| Figure 11(a) | 6.7 GeV | 0.30±0.15 | 1.50±0.10 | 2.70±0.20 | 0.03±0.01 | 0.10±0.02 | 1.00±0.04 | 0.368 | 0.231±0.046 | 0.273±0.019 |
| Figure 11(b) | 11.2 GeV | 0.40±0.10 | 1.85±0.07 | 3.30±0.20 | 0.10±0.02 | 0.08±0.01 | 1.20±0.02 | 0.594 | 0.169±0.035 | 0.279±0.008 |
| Figure 11(c) | 19.4 GeV | 0.40±0.05 | 2.35±0.05 | 3.30±0.10 | 0.13±0.02 | 0.10±0.02 | 1.30±0.04 | 0.422 | 0.131±0.028 | 0.254±0.014 |
| Figure 11(d) | 23.7 GeV | 0.46±0.03 | 2.65±0.05 | 3.80±0.07 | 0.15±0.02 | 0.10±0.02 | 1.35±0.04 | 0.749 | 0.121±0.026 | 0.252±0.013 |
| Figure 11(e) | 38.7 GeV | 0.60±0.08 | 3.30±0.05 | 4.80±0.05 | 0.15±0.02 | 0.04±0.02 | 1.70±0.04 | 0.296 | 0.102±0.022 | 0.321±0.012 |

Table 7. Values of intercept, slope, and $\chi^2/dof$ for the linear relations $y_T - C$ in Figures 12(a), 12(b), and 12(d), for the linear relations $y_{T,P,C} - C$ in Figures 12(c) and 12(e), and for the linear relations $y_{T,P,C} - \sqrt{s_{NN}}$ in Figure 12(f).

| Figure | $\sqrt{s_{NN}}$ or Type | Intercept Value | Error | Slope Value | Error | $\chi^2/dof$ |
|---|---|---|---|---|---|---|
| Figure 12(a) | 19.6 GeV | -1.119 | 0.028 | 0.003 | 0.001 | 0.035 |
|  | 62.4 GeV | -1.691 | 0.044 | 0.001 | 0.001 | 0.085 |
|  | 130 GeV | -1.927 | 0.022 | -0.006 | 0.001 | 0.013 |
|  | 200 GeV | -2.020 | 0.017 | -0.002 | 0.001 | 0.005 |
| Figure 12(b) | 22.4 GeV | -1.167 | 0.019 | 0.003 | 0.001 | 0.032 |
|  | 62.4 GeV | -1.632 | 0.045 | 0.006 | 0.001 | 0.120 |
|  | 200 GeV | -2.319 | 0.014 | 0.002 | 0.001 | 0.006 |
| Figure 12(c) | T | -4.176 | 0.217 | 0.001 | 0.004 | 0.120 |
|  | P | 0.690 | 0.082 | 0.008 | 0.002 | 0.083 |
|  | C | -1.702 | 0.174 | 0.015 | 0.003 | 0.793 |
| Figure 12(d) | 2.76 TeV | -0.487 | 0.058 | 0.001 | 0.004 | 0.149 |
| Figure 12(e) | T | 0.907 | 0.023 | 0.001 | 0.001 | 0.003 |
|  | P | 2.773 | 0.463 | 0.016 | 0.008 | 0.508 |
|  | C | 1.891 | 0.433 | 0.016 | 0.008 | 0.665 |
| Figure 12(f) | T | 0.250 | 0.035 | 0.009 | 0.002 | 0.050 |
|  | P | 2.182 | 0.182 | 0.067 | 0.006 | 0.169 |
|  | C | 1.296 | 0.103 | 0.053 | 0.004 | 0.122 |

Table 8. Values of intercept, slope, and $\chi^2/dof$ for the linear relations $\sigma - C$ in Figures 13(a)-13(e) and for the linear relation $\sigma - \sqrt{s_{NN}}$ in Figure 13(f).

| Figure | $\sqrt{s_{NN}}$ or $E_{beam}$ | Intercept Value | Error | Slope Value | Error | $\chi^2/dof$ |
|---|---|---|---|---|---|---|
| Figure 13(a) | 19.6 GeV | 1.625 | 0.017 | 0.008 | 0.001 | 0.021 |
|  | 62.4 GeV | 1.935 | 0.018 | 0.014 | 0.001 | 0.022 |
|  | 130 GeV | 2.262 | 0.016 | 0.009 | 0.001 | 0.018 |
|  | 200 GeV | 2.423 | 0.026 | 0.008 | 0.001 | 0.029 |
| Figure 13(b) | 22.4 GeV | 1.686 | 0.026 | 0.008 | 0.001 | 0.072 |
|  | 62.4 GeV | 1.987 | 0.012 | 0.011 | 0.001 | 0.015 |
|  | 200 GeV | 2.519 | 0.015 | 0.009 | 0.001 | 0.015 |
| Figure 13(c) | 0.2 TeV | 3.144 | 0.149 | -0.008 | 0.003 | 0.315 |
| Figure 13(d) | 2.76 TeV | 3.609 | 0.009 | 0.001 | 0.001 | 0.001 |
| Figure 13(e) | 10.6$A$ GeV | 0.757 | 0.055 | 0.004 | 0.001 | 0.094 |
| Figure 13(f) | 6.7-38.7 GeV | 0.955 | 0.049 | 0.019 | 0.003 | 0.130 |

Table 9. Values of intercept, slope, and $\chi^2/dof$ for the linear relations $c_s^2 - C$ in Figures 14(a)-14(g) and for the linear relations $c_s^2 - \sqrt{s_{NN}}$ in Figure 14(h).

| Figure | $\sqrt{s_{NN}}$ or Type | Intercept Value | Intercept Error | Slope Value | Slope Error | $\chi^2/dof$ |
|---|---|---|---|---|---|---|
| Figure 14(a) | 19.6 GeV | 0.130 | 0.028 | 0 | 0 | 0 |
| | 62.4 GeV | 0.088 | 0.019 | 0 | 0 | 0 |
| | 130 GeV | 0.073 | 0.016 | 0 | 0 | 0 |
| | 200 GeV | 0.066 | 0.014 | 0 | 0 | 0 |
| Figure 14(b) | 19.6 GeV | 0.364 | 0.004 | 0.003 | 0.001 | 0.019 |
| | 62.4 GeV | 0.354 | 0.005 | 0.004 | 0.001 | 0.030 |
| | 130 GeV | 0.386 | 0.004 | 0.002 | 0.001 | 0.022 |
| | 200 GeV | 0.398 | 0.006 | 0.002 | 0.001 | 0.042 |
| Figure 14(c) | 22.4 GeV | 0.123 | 0.026 | 0 | 0 | 0 |
| | 62.4 GeV | 0.088 | 0.019 | 0 | 0 | 0 |
| | 200 GeV | 0.066 | 0.014 | 0 | 0 | 0 |
| Figure 14(d) | 22.4 GeV | 0.374 | 0.009 | 0.003 | 0.001 | 0.115 |
| | 62.4 GeV | 0.368 | 0.004 | 0.003 | 0.001 | 0.025 |
| | 200 GeV | 0.421 | 0.004 | 0.002 | 0.001 | 0.021 |
| Figure 14(e) | T | 0.066 | 0.014 | 0 | 0 | 0 |
| | C | 0.553 | 0.033 | -0.002 | 0.001 | 0.478 |
| Figure 14(f) | T | 0.042 | 0.009 | 0 | 0 | 0 |
| | C | 0.500 | 0.001 | 0.001 | 0.001 | 0.001 |
| Figure 14(g) | T | 0.310 | 0.056 | 0 | 0 | 0 |
| | C | 0.226 | 0.031 | 0.002 | 0.001 | 0.159 |
| Figure 14(h) | T | 0.223 | 0.022 | -0.004 | 0.001 | 0.170 |
| | C | 0.255 | 0.023 | 0.001 | 0.001 | 0.294 |

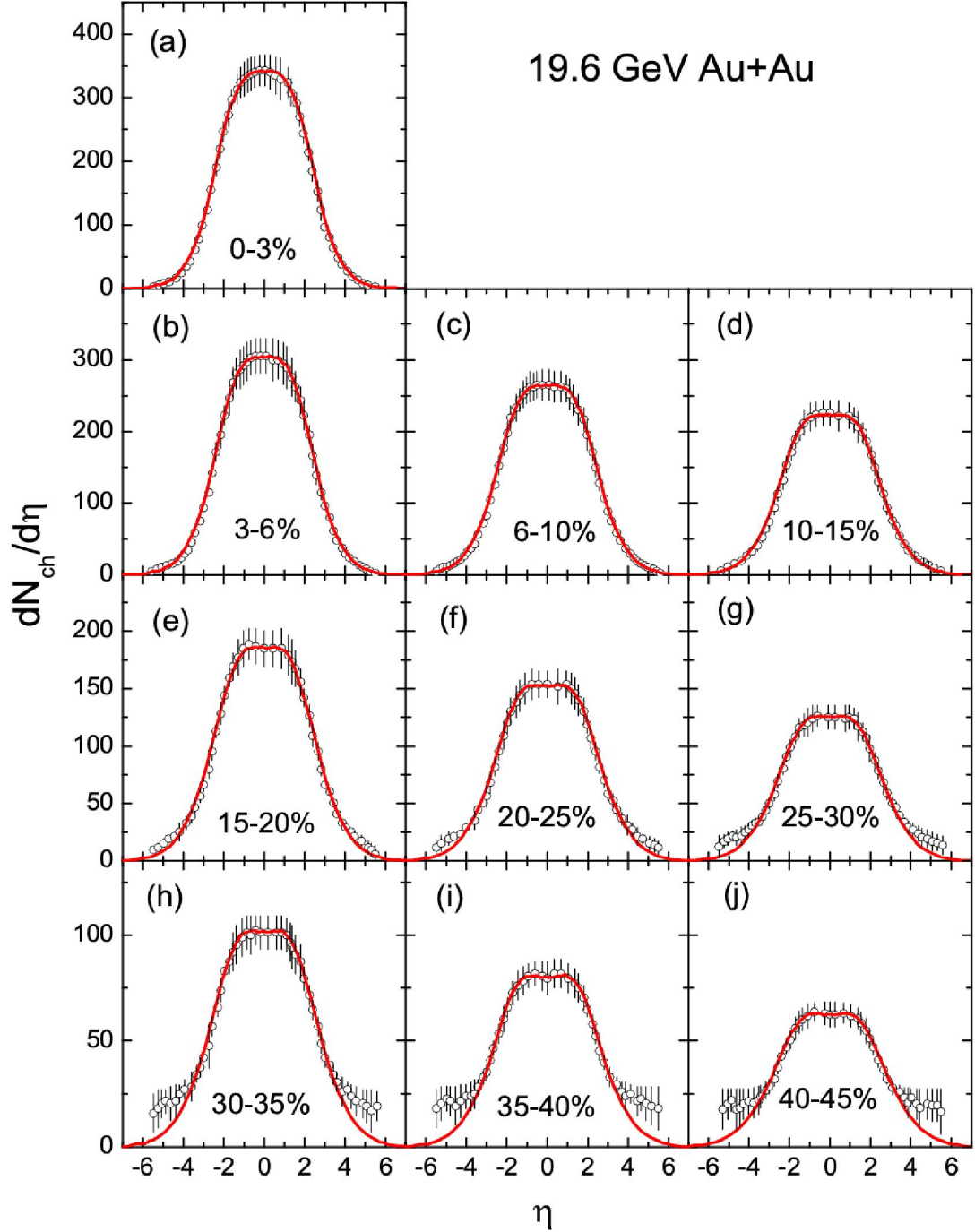

Figure 1: Pseudorapidity distributions of charged particles produced in Au-Au collisions at $\sqrt{s_{NN}} = 19.6$ GeV with different centrality percentages. The circles and curves represent the experimental data quoted in reference [22] and our calculated results respectively. In the calculation, spectator contributions in Figures 1(e)-1(j) are not counted, and there is no spectator being expected in Figures 1(a)-1(d).

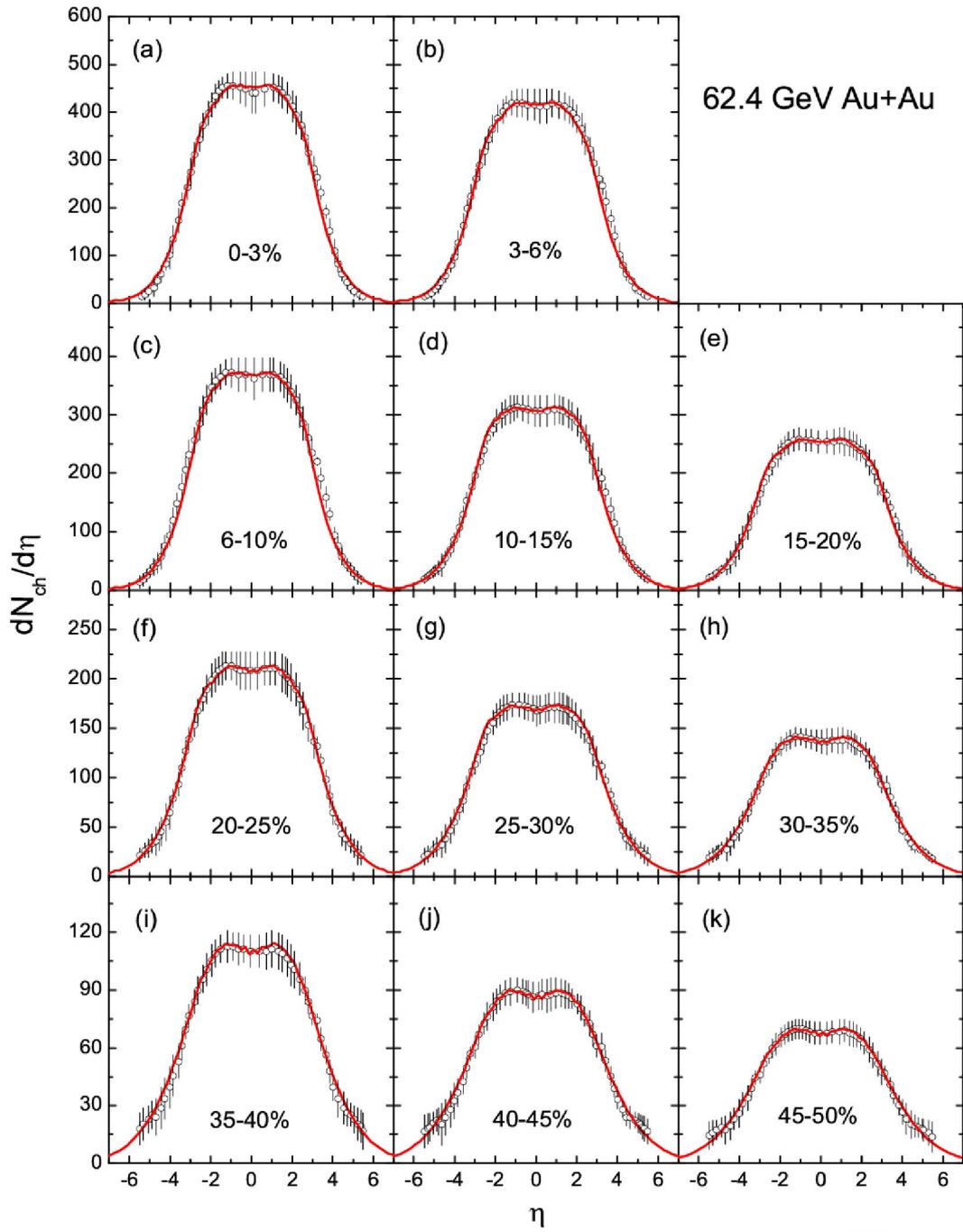

Figure 2: The same as Figure 1, but showing the results of Au-Au collisions at $\sqrt{s_{NN}} = 62.4$ GeV.

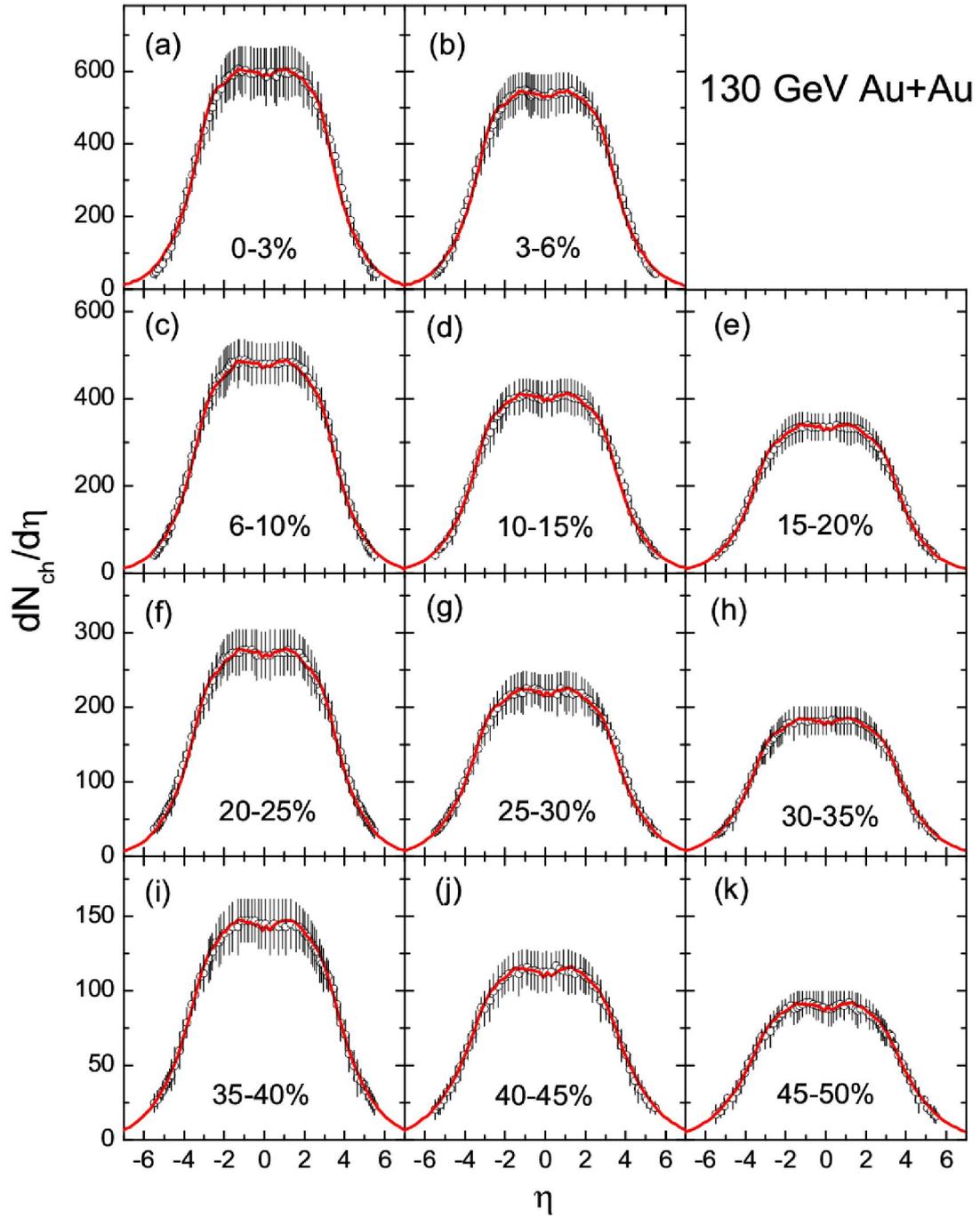

Figure 3: The same as Figure 1, but showing the results of Au-Au collisions at $\sqrt{s_{NN}} = 130$ GeV.

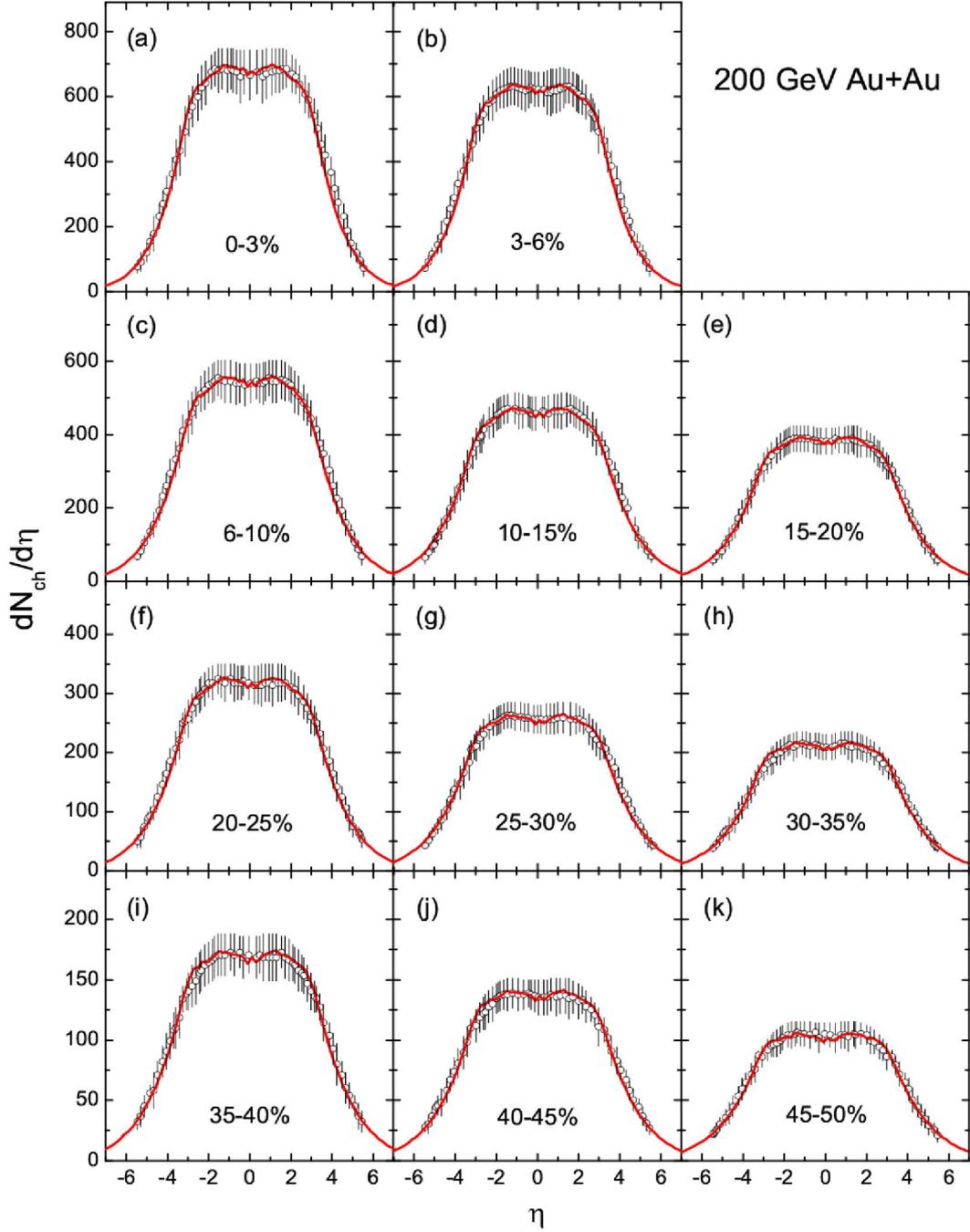

Figure 4: The same as Figure 1, but showing the results of Au-Au collisions at $\sqrt{s_{NN}} = 200$ GeV.

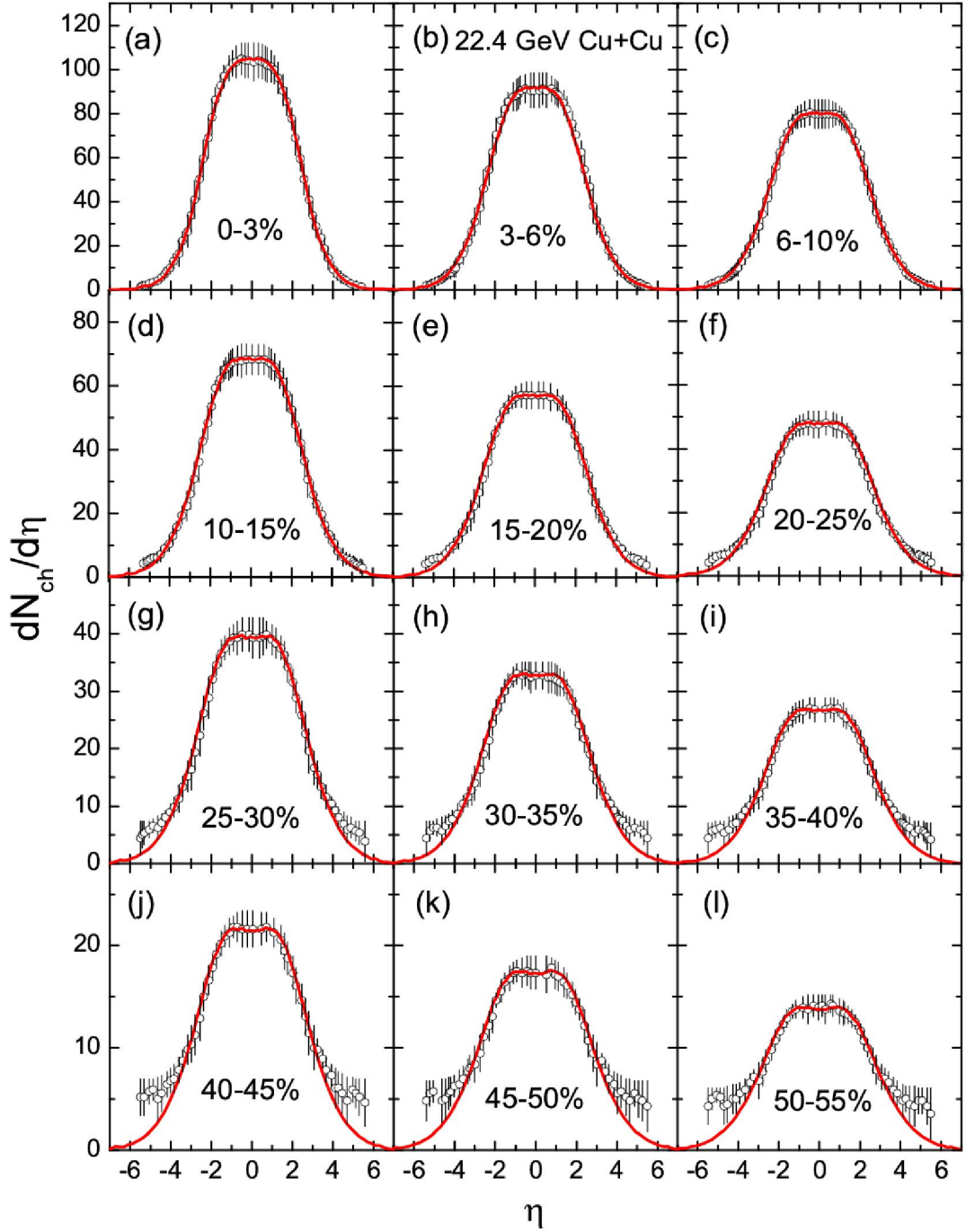

Figure 5: The same as Figure 1, but showing the results of Cu-Cu collisions at $\sqrt{s_{NN}} = 22.4$ GeV. In the calculation, spectator contributions in Figures 5(d)-5(l) are not counted, and there is no spectator being expected in Figures 5(a)-5(c).

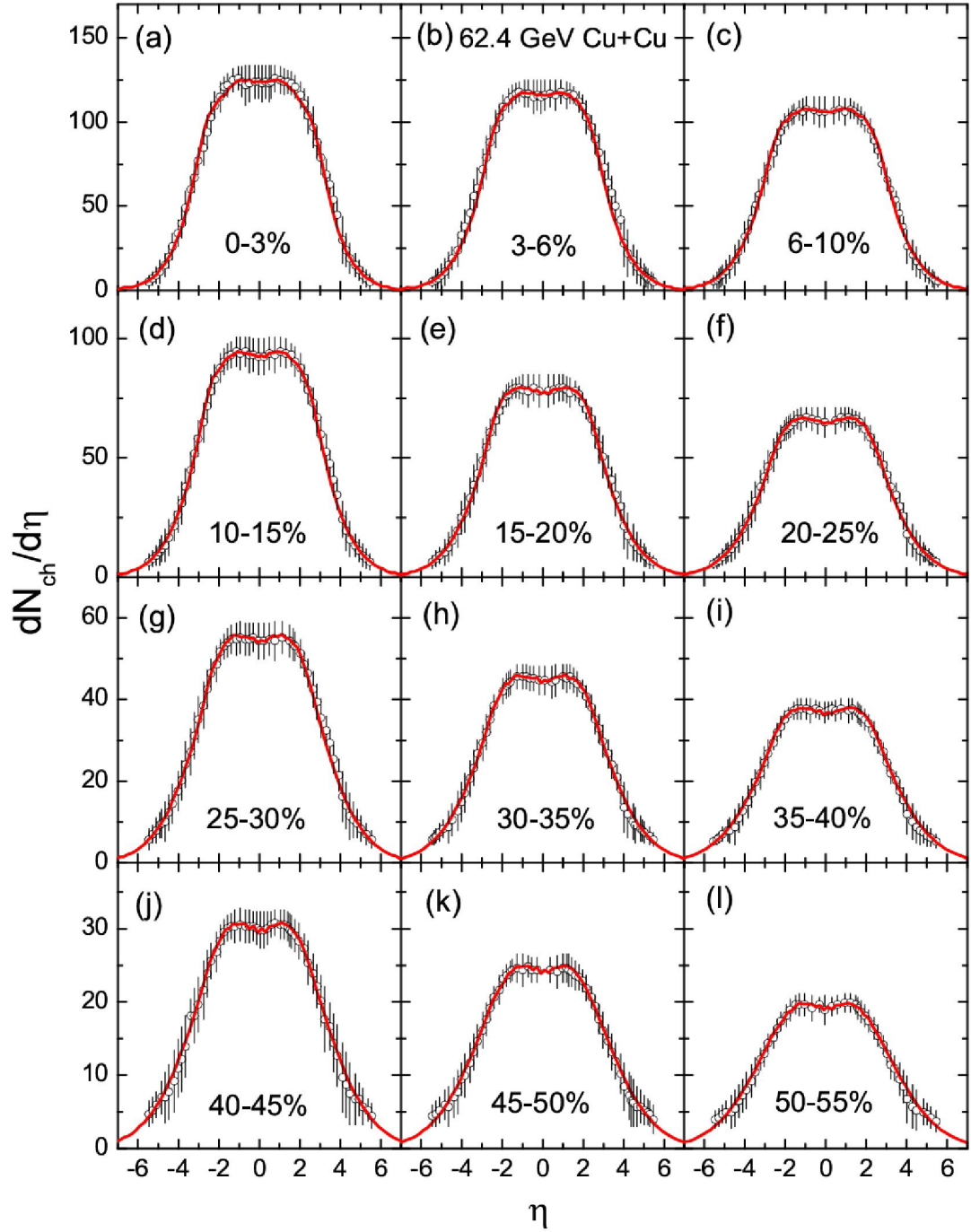

Figure 6: The same as Figure 1, but showing the results of Cu-Cu collisions at $\sqrt{s_{NN}} = 62.4$ GeV.

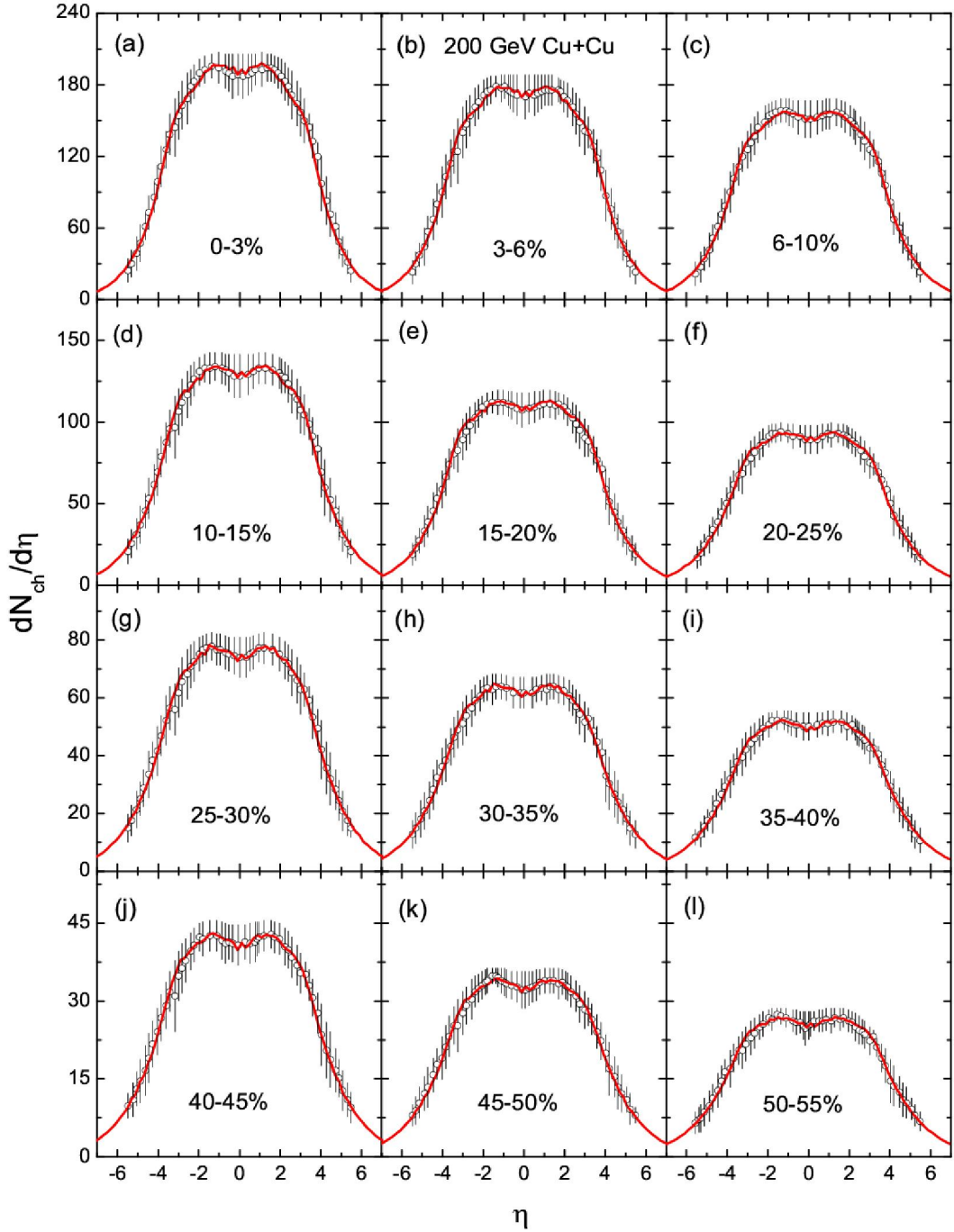

Figure 7: The same as Figure 1, but showing the results of Cu-Cu collisions at $\sqrt{s_{NN}} = 200$ GeV.

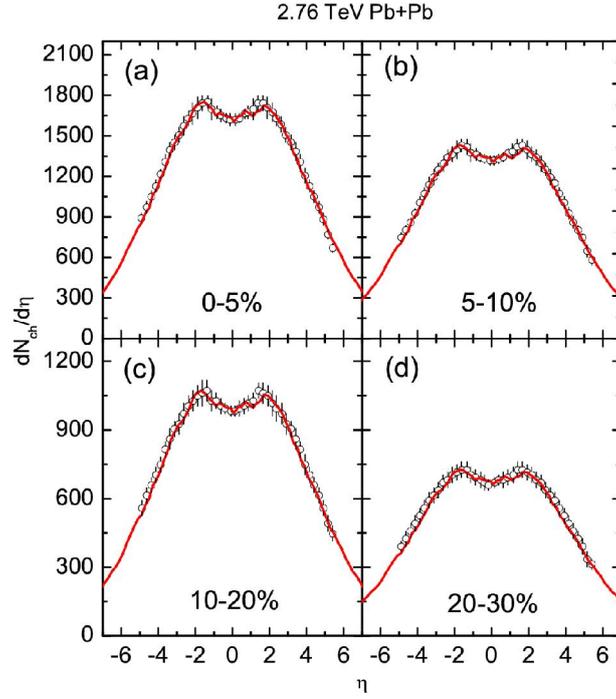

Figure 8: The same as Figure 1, but showing the results of Pb-Pb collisions at $\sqrt{s_{NN}} = 2.76$ TeV, where the data are quoted in reference [26].

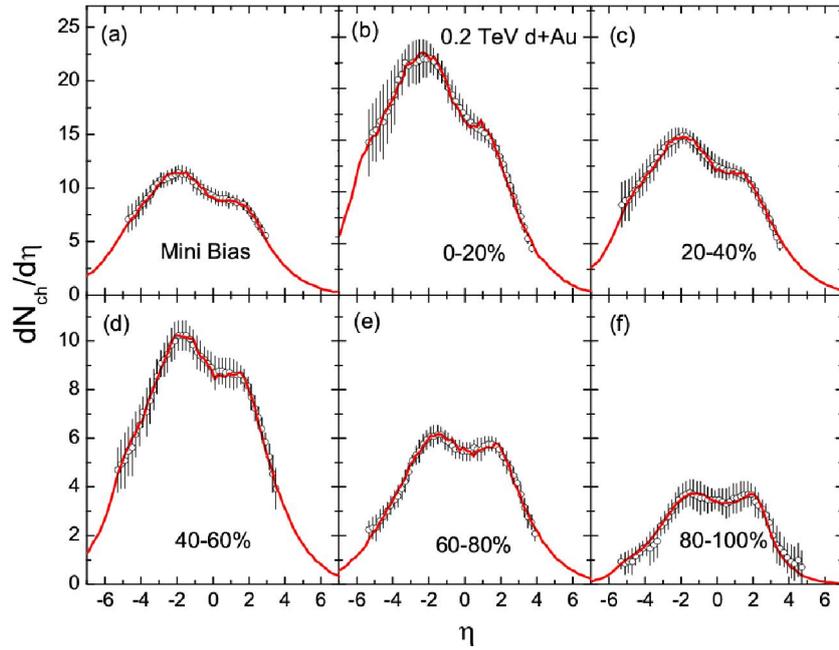

Figure 9: The same as Figure 1, but showing the results of $d$-Au collisions at $\sqrt{s_{NN}} = 0.2$ TeV, where the data are quoted in references [23, 24] [Figure 9(a)] and [25] [Figures 9(b)-9(f)].

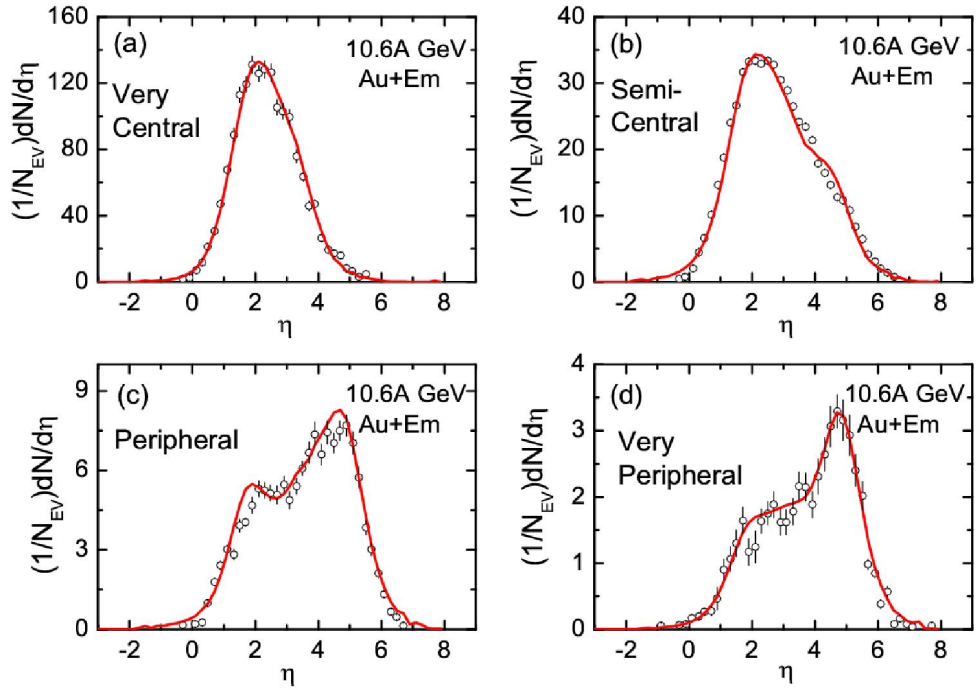

Figure 10: The same as Figure 1, but showing the results of Au-Em collisions at $E_{\text{beam}} = 10.6A$ GeV, where the data are quoted in reference [27].

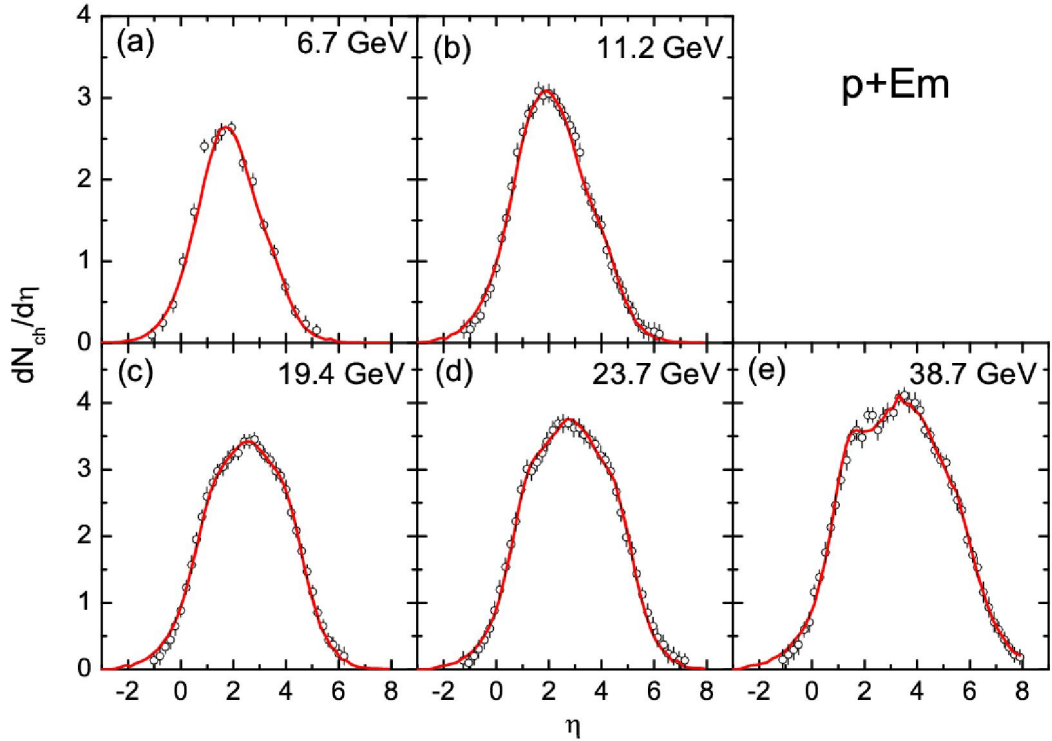

Figure 11: The same as Figure 1, but showing the results of $p$-Em collisions at different energies, where the data are quoted in references [28, 29].

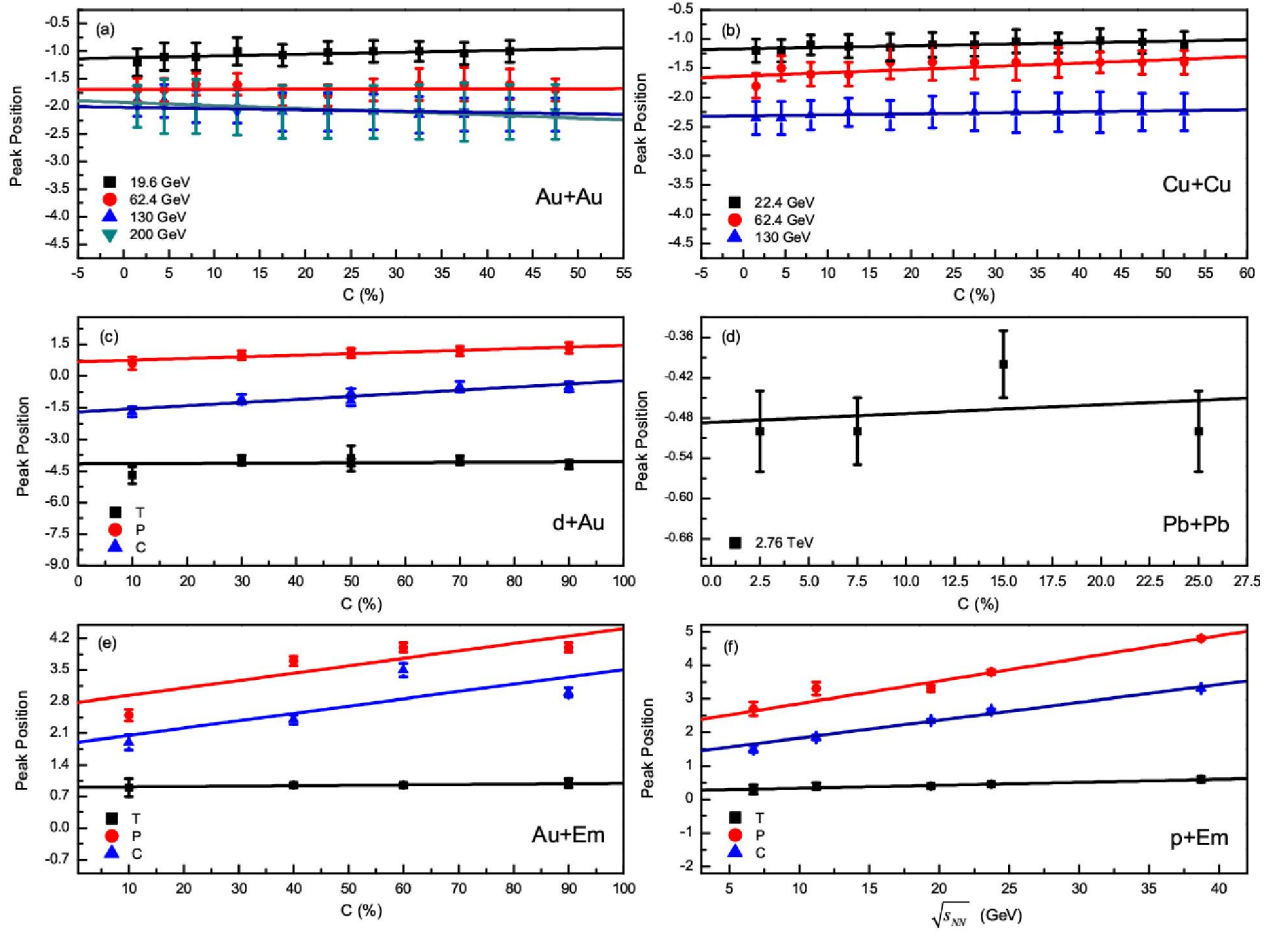

Figure 12: Dependences of peak positions ($y_T$, $y_C$, $y_P$) of rapidity distributions in Au-Au, Cu-Cu, $d$-Au, Pb-Pb, Au-Em, and $p$-Em collisions on $C$ or $\sqrt{s_{NN}}$, where for symmetric collisions $y_C$ and $y_P$ are not presented due to $y_C = 0$ and $y_P = -y_T$. The symbols are quoted in Tables 1-6, and the lines are our fitted results.

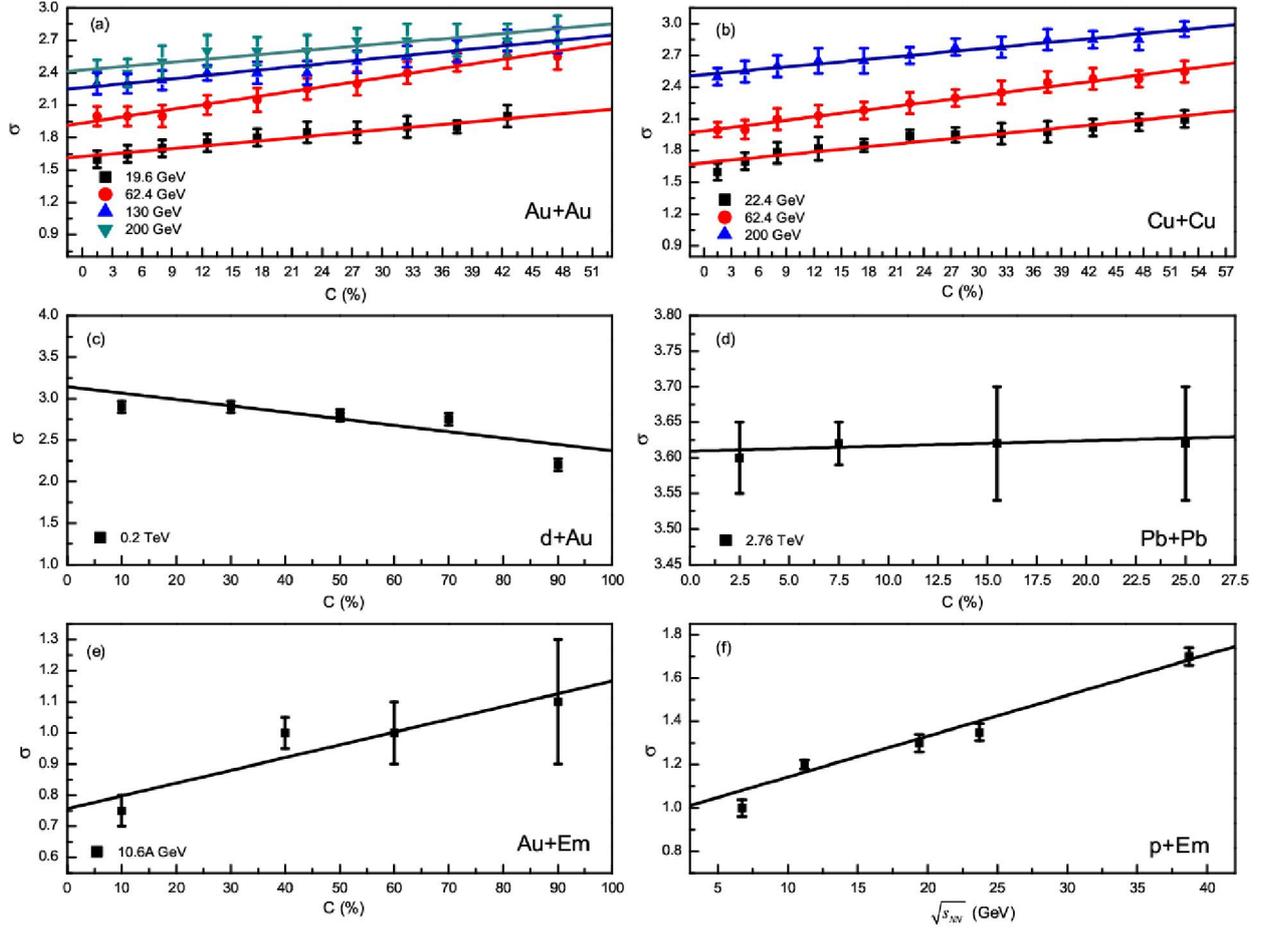

Figure 13: Dependences of $\sigma$ in central sources in Au-Au, Cu-Cu, *d*-Au, Pb-Pb, Au-Em, and *p*-Em collisions on *C* or $\sqrt{s_{NN}}$. The symbols are quoted in Tables 1-6, and the lines are our fitted results.

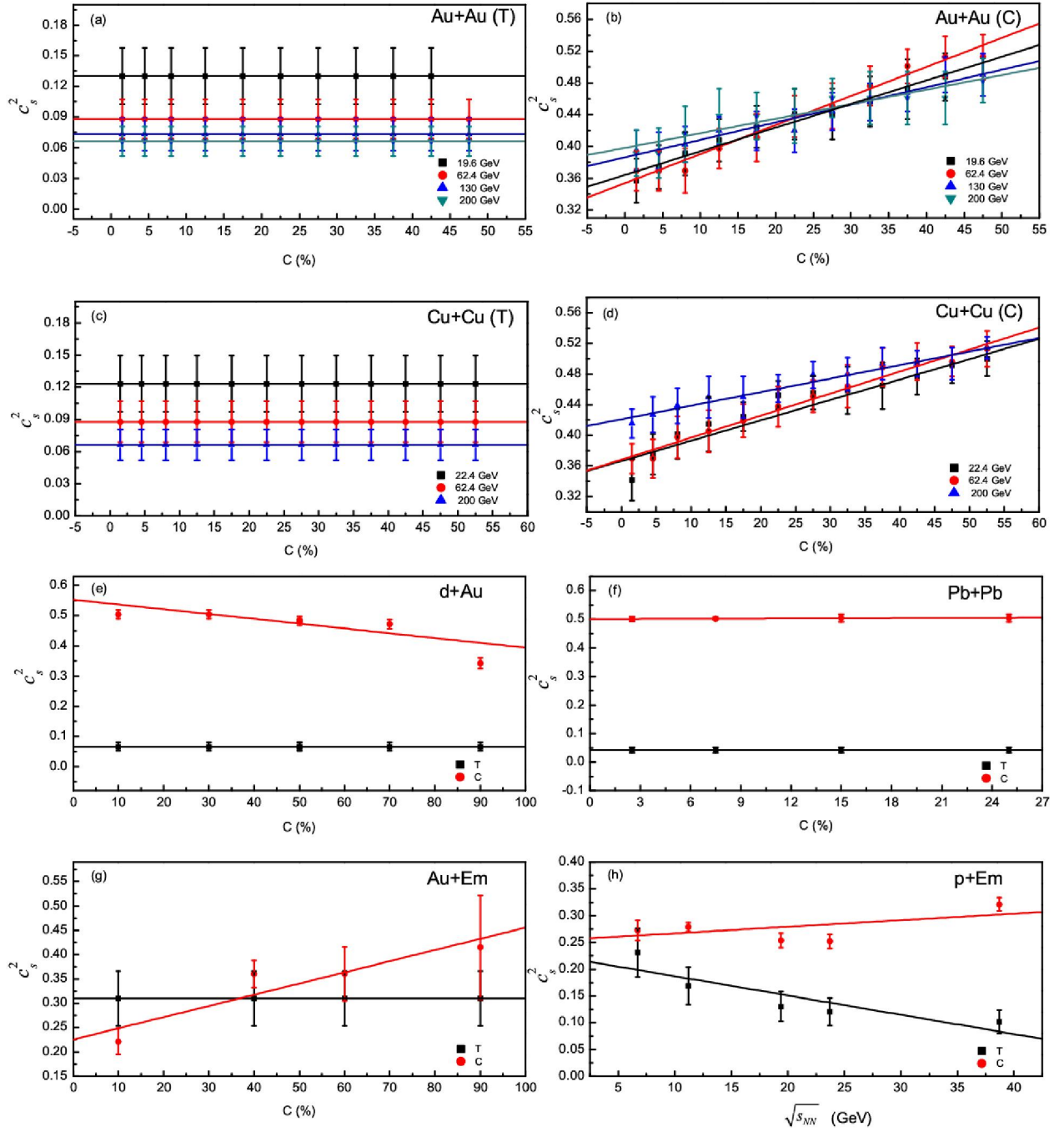

Figure 14: Dependences of $c_s^2$ for target [T or projectile (P)] and central sources (C) in Au-Au, Cu-Cu, $d$-Au, Pb-Pb, Au-Em, and $p$-Em collisions on $C$ or $\sqrt{s_{NN}}$. The symbols are quoted in Tables 1-6, and the lines are our fitted results.